\documentclass[12pt,psf,epsf]{article}
\usepackage[centertags]{amsmath}
\usepackage{amssymb}
\usepackage{graphicx}
\usepackage{epsfig}
\usepackage{ulem}
\usepackage[english]{babel}
\usepackage{array}
\usepackage{amsthm}
\usepackage{latexsym}
\usepackage[mathcal]{euscript}
\pdfoutput=1
\usepackage{epsfig}
\usepackage{jheppub}
\usepackage{mathdots}
\usepackage{MnSymbol}
\usepackage{multirow}

\newcommand{\nn}{\nonumber}
\newcommand{\be}{\begin{equation}}
\newcommand{\ee}{\end{equation}}
\newcommand{\bea}{\begin{eqnarray}}
\newcommand{\eea}{\end{eqnarray}}

\newcommand{\e}{\mathrm{e}}
\newcommand{\ZZ}{\mathbb{Z}}

\newcommand{\RR}{\mathbb{R}}
\newcommand{\dd}{\partial}
\newcommand{\sgn}{\text{sgn}}

\newcommand{\mJ}{\mathcal{J}}

\newcommand{\mL}{\mathcal{L}}

\renewcommand{\Re}{\text{Re}}
\renewcommand{\Im}{\text{Im}}

\DeclareMathOperator{\Pf}{Pf}

\DeclareMathOperator{\Tr}{Tr}

\newcommand{\fg}{\mathfrak{g}}

\definecolor{dgreen}{rgb}{0.,0.6,0.}

\title{Spectral form factor \\ in the double-scaled SYK model}
\author[a]{Mikhail Khramtsov,}
\author[b,c]{Elena Lanina}

\affiliation[a]{Steklov Mathematical Institute, Russian Academy of Sciences,\\Gubkina str. 8, 119991, Moscow, Russia}
\affiliation[b]{Moscow Institute of Physics and Technology,\\Institutskii per, 9, 141700, Dolgoprudny, Russia}
\affiliation[c]{Institute for Theoretical and Experimental Physics,\\B. Cheremushkinskaya, 25, 117218, Moscow, Russia}
\emailAdd{khramtsov@mi-ras.ru}
\emailAdd{lanina.en@phystech.edu}

\abstract{In this note we study the spectral form factor in the SYK model in large $q$ limit at infinite temperature. We construct analytic solutions for the saddle point equations that describe the slope and the ramp regions of the spectral form factor time dependence. These saddle points are obtained by taking different approaches to the large $q$ limit: the slope region is described by a replica-diagonal solution and the ramp region is described by a replica-nondiagonal solution. We find that the onset of the ramp behavior happens at the Thouless time of order $q \log q$. We also evaluate the one-loop corrections to the slope and ramp solutions for late times, and study the transition from the slope to the ramp. We show this transition is accompanied by the breakdown of the perturbative $1/q$ expansion, and that the Thouless time is defined by the consistency of extrapolation of this expansion to late times. }

\begin{document}
\maketitle

\section{Introduction}

The SYK model \cite{Sachdev92,Kitaev,MScomments,Kitaev17} has proved to be a very useful tool for studies of quantum gravity in $2$d \cite{Jensen16,Maldacena16,Engelsoy16,Jevicki16,Cotler16,Maldacena18,Harlow18,Saad18,Goel18,SSS,Saad19,Garcia-Garcia19,Gao19,Garcia-Garcia20} and general properties of chaotic quantum systems \cite{Garcia-Garcia16,Cotler16,Garcia-Garcia18,Garcia-Garcia19,Saad19,Qi18,Altland17,Gharibyan18,Sonner17,Hunter-Jones17,Jia19,delCampo17,Nayak19,Altland20,Lau18,Lucas19,Choi20,Roberts18,Winer20,You16,Li17,Cardella19}. The SYK model played a big part in inspiring the recent developments such as deep connections between the Jackiw-Teitelboim gravity and double-scaled matrix models \cite{SSS,Arefeva192,Stanford19,Johnson19,Okuyama19} and steps towards resolution of the black hole information paradox \cite{Penington191,Penington192,Almheiri191,Almheiri192,Almheiri194}. 

One of the reasons why SYK-like models are special is that they can be described as a collective field path integral where the large $N$ limit is explicitly semiclassical. On the other hand, SYK is not a true quantum system, but is an ensemble of quantum systems. The interplay between the ensemble averaging and semiclassical nature of the large $N$ expansion raises questions about its structure, in particular about the replica trick and the difference between quenched and annealed averaging \cite{Gur-Ari18,AKTV,Wang18,Arefeva19}. The disorder averaging was shown to be of great benefit for studies of fine-grained quantum chaos in SYK \cite{Cotler16,Garcia-Garcia16,Gharibyan18,Saad18,Sonner17,delCampo17,Jia19}, because it provides a mechanism to emphasize the universal RMT behavior which is built into the theory. The work \cite{Saad18} also showed that some of this universality can be described neatly in terms of the semiclassical formulation as a nontrivial large $N$ saddle point. Meanwhile, in the work \cite{AKTV,AKV} we have demonstrated that there are other nontrivial large $N$ saddle points in the Euclidean partition function, and it is not clear what is their physical meaning. Some of these saddle points are the replica-nondiagonal solutions, which exist for integer number of SYK copies (replicas), even if they do not interact. They break the factorization, similarly to the replica wormholes in gravity \cite{Almheiri194,Penington192}.

Motivated by this question, in the present work we study the spectral form factor in the SYK model. Spectral form factor for a quantum mechanical system at finite temperature $\beta^{-1}$ is defined as 
\be
\frac{Z(\beta + iT) Z (\beta - iT)}{Z(\beta)^2} = \frac{1}{Z(\beta)^2}\Tr\ \e^{-\beta H -i H T } \Tr\ \e^{-\beta H + i H T} =\frac{1}{Z(\beta)^2} \sum_{n, m} \e^{-\beta(E_m + E_n)} \e^{i T(E_m - E_n)}\,. \label{spectral form factor}
\ee
It was studied numerically in the SYK model in \cite{Cotler16,Gharibyan18,Gur-Ari18,Sonner17,delCampo17} by exact diagonalization of the Hamiltonian for some finite $N$ and averaging over the disorder afterwards. We are interested in the role of large $N$ saddle points, so we study this quantity in terms of the collective field path integral, following \cite{Saad18}, where the saddle point solutions were constructed numerically. Our task is to find all large $N$ saddle points that contribute to the spectral form factor in the slope and ramp regimes. To make the problem analytically tractable, we study the SYK model in the large $q$, or double-scaled limit \cite{MScomments,Cotler16}. This limit has been utilized extensively to simplify the computations in SYK and related models \cite{Maldacena18,Gao19,Roberts18,Gharibyan18,Qi18,Choi19,Choi20,Schroder,Eberlein17,Bhattacharya17,Berkooz18,Berkooz20,Tarnopolsky18,Jiang19,Streicher19,Nosaka19,Das20}.

In this paper we construct analytic solutions of the saddle point equations in the double-scaled SYK which describe the slope and ramp regimes of the spectral form factor at $\beta= 0$, and evaluate the one-loop corrections to these saddle points in the late time limit. We pay special attention to applicability of the $1/q$-expansion at late times. Considering the consistency conditions for solutions allow us to estimate the time of the onset of the ramp behavior, and the time of the transition from the ramp to the slope. 

The paper is organized as follows. In the section \ref{sec:Setup} we discuss the collective field path integral for the spectral form factor in the SYK model and discuss the ways of taking the large $q$ limit that would be appropriate for our purposes. In section \ref{sec:Slope}, we focus on the disconnected part of the spectral form factor. We construct the analytic saddle point solutions in the large $q$ limit, verify their validity at late times, compute their contribution to the spectral form factor and take into account the 1-loop correction. In section \ref{sec:Ramp} we give the same treatment to the connected spectral form factor, obtaining the analytic replica-nondiagonal solution and verifying that it describes the ramp region. In section \ref{sec:Transition} we discuss the different notable time scales where the dynamics of the spectral form factor changes. Finally, in the section \ref{sec:Discussion} we discuss the results and open questions. 

\section{Setup}
\label{sec:Setup}

\subsection{Spectral form factor in the SYK model}

We study the SYK model as defined by the Hamiltonian \cite{Sachdev92,Kitaev}:
\be
H = i^{q/2} \sum_{i_1< i_2< \dots < i_q=1}^N j_{i_1 i_2\dots i_q} \psi_{i_1} \psi_{i_2} \dots \psi_{i_q}\,. \label{Hsyk} 
\ee
Here $\psi_i$ are the Majorana fermions, and $ j_{i_1\dots i_q}$ are totally antisymmetric couplings randomized via the Gaussian distribution: 
\be
P(j_{i_1\dots i_q}) = \sqrt{\frac{q N^{q-1}}{2^q (q-1)!\pi \mJ^2}}
\ \e^{-\frac{q N^{q-1} j_{i_1 \dots i_q}^2}{2^q (q-1)! \mJ^2}}\,. \label{Gaussian}
\ee
The distribution (\ref{Gaussian}) enforces the following rules for disorder averaging: 
\be
\langle j_{i_1 \dots i_q} \rangle = 0\,, \qquad \langle j_{i_1 \dots i_q}j_{i_1 \dots i_q} \rangle =  \frac{2^{q-1} (q-1)! \mJ^2}{q N^{q-1}} \quad \text{(no sum)}\,. \label{disorder2point}
\ee
Note that in studies beyond the large $q$ limit instead of $\mJ$ one typically uses 
\be
J^2 = \frac{2^{q-1}}{q} \mJ^2\,. \label{normalJ}
\ee
Since the SYK model has a random ensemble of couplings, we are going to study the averaged variant of the spectral form factor (\ref{spectral form factor})\footnote{Note that we take separate average for nominator and denominator. This is the annealed average, which, generally speaking, gives different result compared to the quenched average, for which the entire fraction is averaged, necessitating the replica trick \cite{AKTV}. However, it is known that at least under the applicability of large $N$ expansion, there are no replica-nondiagonal phases in quenched quantities in the full SYK model \cite{AKTV,Gur-Ari18,Wang18}, so we expect that the annealed average coincides with the quenched average for the leading order in $1/N$.}:
\be
S(\beta, T) = \frac{1}{\langle Z(\beta)^2\rangle} \langle Z(\beta + iT) Z(\beta - iT) \rangle\,. \label{S}
\ee
In the present work we focus on the case $\beta = 0$, so more specifically we study the quantity
\be
S(T) = \frac{ \langle Z(iT) Z(- iT) \rangle}{\langle Z(0)^2 \rangle} \,, \label{S(T)}
\ee
where $Z(0) = 2^{\frac{N}{2}}$ is the SYK partition function at infinite temperature. One can decompose the full spectral form factor $S(T)$ into connected and disconnected parts as follows:
\be
S(T) = \frac{| \langle Z(iT) \rangle |^2}{\langle Z(0)^2 \rangle} + K(T)\,,
\ee
where $K(T)$ is the connected spectral form factor:
\be
K(T) = \frac{\langle Z(iT) Z(-iT) \rangle}{\langle Z(0)^2 \rangle} - \frac{\langle Z(iT) \rangle \langle Z(-iT) \rangle}{\langle Z(0)^2 \rangle} \,.
\ee
\begin{figure}[t]
	\centering
	\includegraphics[scale=0.6]{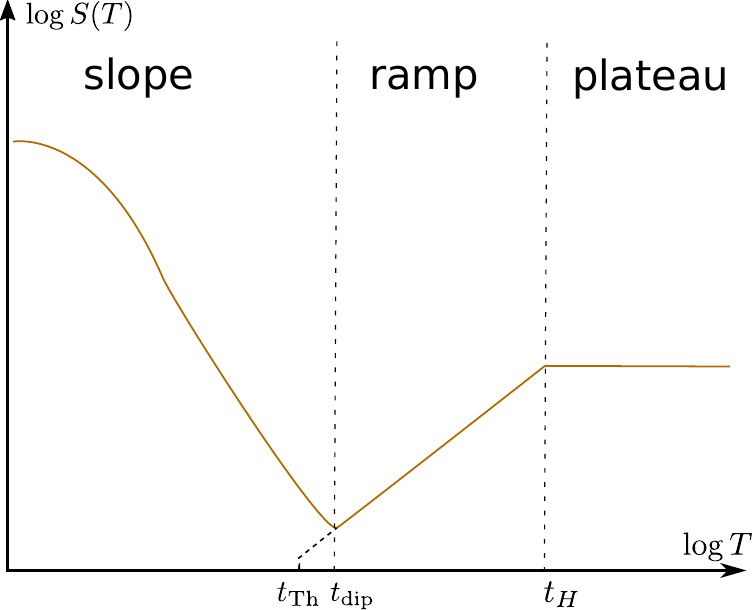}
	\caption{A sketch of the spectral form factor time dependence.}
	\label{fig:SketchSFF}
\end{figure}
The sketch of the time dependence of the spectral form factor in SYK is shown on Fig. \ref{fig:SketchSFF}, as demonstrated by the numerical results in \cite{Cotler16,Gharibyan18,Saad18}. In the chaotic systems the disconnected part of the spectral form factor dominates at early times, governing what is known as \textit{slope} region. The connected part of the spectral form factor dominates at late times, governing the so-called \textit{ramp} and \textit{plateau} regions \cite{Cotler16,Garcia-Garcia16,Gharibyan18,Saad18}. Let us review the basic properties of these regimes. 

\paragraph{The slope regime} is defined by the disconnected part of the spectral form factor as a square of the Fourier transform of the spectral density. It is also self-averaging in the disorder ensemble. 

\paragraph{The ramp regime} is the linear growth of the connected spectral form factor. It starts from what is known as ergodic, or Thouless time \cite{Altland17,Altland20,Nosaka18,Saad18,Gharibyan18,Garcia-Garcia16,Sonner17} $T_{\text{Th}}$, starts dominating over the ramp at so-called dip time $T_{\text{dip}}$ \cite{Cotler16,Gharibyan18} and continues until the Heisenberg time $T_H$ proportional to the inverse mean level spacing. The ramp is not self-averaging, but the disorder average shows the behavior which agrees with the RMT universality \cite{Cotler16}. 

\paragraph{The plateau regime} is the approximately constant behavior after $T_H$. It is also not self-averaging. As discussed in recent work \cite{Saad18,Altland20,Saad19,SSS}, it is determined by the very small nonperturbative effects which in SYK terms scale like $\e^{-\e^N}$, corresponding to nonperturbative effects in RMT. The plateau regime is inaccessible to our analysis and is beyond the scope of the present work.

Now let us discuss the expression for $S(T)$ in the SYK variables. We have:
\bea
&& Z( iT) = \Tr \e^{- i T H} = \int D\psi \exp\left[ i \int_0^T dt \left(\frac{i}{2} \psi_i \dd_t \psi_i - i^{q/2}\!\!\!\!\! \sum_{i_1< \dots < i_q=1}^N \!\!\!\!\! j_{i_1 i_2\dots i_q} \psi_{i_1} \dots \psi_{i_q}\right)\right];  \label{Z(iT)}\\
&& Z(-iT) =\int D\psi \exp\left[ i \int_0^T dt \left(\frac{i}{2} \psi_i \dd_t \psi_i + (-i)^{q/2} \!\!\!\!\! \sum_{i_1<  \dots < i_q=1}^N \!\!\!\!\! j_{i_1 i_2\dots i_q} \psi_{i_1} \dots \psi_{i_q}\right)\right]\,.  \label{Z(-iT)}
\eea
The trace implies the anti-periodic boundary conditions for the fermions in the path integral: 
\be
\psi_i (T) = -\psi_i(0) \label{anti-periodicity}\,.
\ee
After averaging over the disorder, one can rewrite $S(T)$ as a path integral over collective bilocal fields $G$ and $\Sigma$ \cite{Saad18}, analogously to the replica partition function for the Euclidean SYK at finite temperature \cite{Kitaev,MScomments,Kitaev17,AKTV}. We sketch this derivation in Appendix \ref{sec:ReplicasDeriv}. The resulting expression has the form 
\be
S(T) =\frac{1}{\langle Z(0)^2 \rangle} \int DG_{\alpha\beta} D \Sigma_{\alpha\beta} \e^{-N I[G, \Sigma]}\,, \label{S(T)-bilocal}
\ee
where the action has the form \cite{Saad18}:
\bea \label{action}
I[G, \Sigma] 
	=&&  -\log \Pf[\delta_{\alpha\beta} \dd_t - \hat{\Sigma}_{\alpha\beta}] + \\&&\nn \frac{1}{2}\int_0^T \int_0^T dt_1 dt_2  \left(\Sigma_{\alpha\beta}(t_1, t_2) G_{\alpha\beta}(t_1, t_2) - \frac{2^{q-1} \mJ^2}{q^2}s_{\alpha\beta} G_{\alpha\beta}(t_1, t_2)^q\right)\,.
\eea
Here the $\alpha,\beta = L,R$ are the indices which enumerate the two copies (replicas) of the SYK system, $\Pf$ denotes the Pfaffian in both time and replica indices, and the hat denotes the integral operator defined by the corresponding bilocal field as its kernel. The $s_{\alpha\beta}$ is a constant matrix with the following elements: 
\be
s_{LL} = s_{RR} = -1\,; \quad s_{LR} = s_{RL} = i^q \,. \label{s}
\ee
The bilocal collective fields $G$ and $\Sigma$ satisfy the antisymmetry constraint: 
\be
G_{\alpha\beta}(t_1, t_2) = -G_{\beta\alpha}(t_2, t_1)\,. \label{antisymmetry}
\ee
The saddle point equations for the action (\ref{action}) read: 
\be \label{EOM}
\begin{aligned}
	&\partial_{t_1} G_{\alpha\beta}(t_1,t_2)-\int dt\Sigma_{\alpha\gamma}(t_1,t)G_{\gamma\beta}(t,t_2)=\delta(t_1-t_2)\delta_{\alpha\beta}, \\
	&\Sigma_{\alpha\beta}(t_1,t_2)=s_{\alpha\beta}\frac{2^{q-1}\mJ^2}{q} G_{\alpha\beta}(t_1,t_2)^{q-1}.
\end{aligned}
\ee
The goal of the present work is to find and study the main solutions of these saddle point equations which determine the dynamics of the spectral form factor, particularly at late times. 

\subsection{The large $q$ limits}
\label{sec:largeQ}

To solve these equations analytically, we will make use of the key simplifying assumption, the limit of large $q$. In order for the full path integral (\ref{S(T)-bilocal}) to remain well-defined in this limit, it is natural to also set $N\to \infty$ and introduce the new fixed parameter $\lambda$:
\be
N \to \infty\,, \quad q \to \infty\,, \quad \lambda = \frac{q^2}{N}\; \text{fixed} \label{DS}
\ee
This is known as double-scaled limit of SYK \cite{Cotler16,MScomments,Gharibyan18}. Our study of the spectral form factor is based on finding the saddle points of the path integral (\ref{S(T)-bilocal}). The constant $\lambda$ plays the role of the new Planck constant. Besides the saddle point contribution, we are also going to be interested in the one-loop corrections, which are the leading-order correction in the expansion in $\lambda$, assuming it is small. 

To find the analytic solutions of the saddle point equations in the large $q$ limit, we will use a generalized version of the approach used previously in the SYK literature \cite{MScomments,Maldacena18,Choi19,Qi18,Streicher19,Berkooz18,Jiang19,Choi20,Qi18,Roberts18,Nosaka19,Das20}.

The basic idea is to expand the bilocal fields $G_{\alpha\beta}$ and $\Sigma_{\alpha\beta}$ in series in $1/q\rightarrow 0$:
\be 
\begin{aligned}\label{1/q-expansion}
	G_{\alpha\beta}&=G_{\alpha\beta}^{(0)}+\frac{1}{q}G_{\alpha\beta}^{(1)}+\frac{1}{q^2}G_{\alpha\beta}^{(2)}+\dots, \\
	\Sigma_{\alpha\beta}&=\Sigma_{\alpha\beta}^{(0)}+\frac{1}{q}\Sigma_{\alpha\beta}^{(1)}+\frac{1}{q^2}\Sigma_{\alpha\beta}^{(2)}+\dots.
\end{aligned}
\ee
First thing to note is that this expansion does not converge for all times. Since we are interested in the behavior of the spectral form factor on arbitrarily late time scales, we would have to extrapolate the results given by the above $1/q$-expansion for late times. So we will proceed as follows: 
\begin{itemize}
	\item[1.] Find the solutions of saddle point equations (\ref{EOM}) for times small enough where the $1/q$-expansion is applicable
	\item[2.] Find an approximate solutions in the late time regime 
	\item[3.] Glue the two regimes together, use the smoothness conditions to establish global existence of solutions.
\end{itemize}
To this end, let us start assuming that the times are small: 
\be
\mJ T \sim 1 \ll q\,,
\ee
so that the expansion (\ref{1/q-expansion}) is applicable. We will expand and solve the equations (\ref{EOM}) up to the first order in $1/q$. In 0th order of $1/q$, we have the system of equations
\begin{equation}\label{zeroOrderSol}
	\begin{aligned}
		\partial_{t_1}G_{LL}^{(0)}&=\partial_{t_1}G_{RR}^{(0)}=\delta(t_1-t_2)\,; \\
		\partial_{t_1}G_{LR}^{(0)}&=\partial_{t_1}G_{RL}^{(0)}=0\,;\\
		\Sigma_{\alpha\beta}^{(0)}&=0\,.
	\end{aligned}
\end{equation}
The first equation for the diagonal components of $G_{\alpha\beta} $ can be solved by the free fermion propagator:
\be
G_{LL}^{(0)}=G_{RR}^{(0)}=G_f(t_1-t_2) = \frac{1}{2}\sgn(t_1-t_2)\,. \label{G_f}
\ee
The second equation for the off-diagonal components can be solved by a constant, taking into account the constraint $G_{\alpha\beta}(t_1, t_2) = -G_{\beta\alpha}(t_2, t_1)$. This constant determines the particular large $q$ limit, which we will have to use to describe different regimes of the spectral form factor. Specifically, we expect the disconnected part of the spectral form factor to be described by a solution that is diagonal in replica indices in all orders of $1/q$, and we expect the connected part of the spectral form factor to be described by a non-diagonal solution. One can show that in order to obtain a non-trivial nondiagonal solution in all orders of $1/q$, the zeroth order of $G$ must be nondiagonal, otherwise the expansion (\ref{1/q-expansion}) will be inconsistent with the equations of motion (\ref{EOM}). 

Considering now the first order in $1/q$ of the expansion for $G$, we get the equations:
\bea
&&	\partial_{t_1}G^{(1)}_{\alpha\beta}(t_1,t_2)-\int dt\Sigma^{(1)}_{\alpha\gamma}(t_1,t)G^{(0)}_{\gamma\beta}(t,t_2)=0\,;\\
&&	\Sigma_{\alpha\beta}(t_1,t_2)=s_{\alpha\beta} \mJ^2 \frac{2^{q-1}}{q} \left(G_{\alpha\beta}^{(0)}+\frac{1}{q}G_{\alpha\beta}^{(1)}+\dots\right)^{q-1}\,.\label{Sigma^1-eq}
\eea
Applying the derivative $\partial_{t_2}$ to the left of the first equation, one gets:
\begin{equation}
	\partial_{t_1}\partial_{t_2}G^{(1)}_{\alpha\beta}(t_1,t_2)=-\Sigma^{(1)}_{\alpha\beta}(t_1,t_2)\,. \label{G^1-eq}
\end{equation} 
Together with the equation (\ref{Sigma^1-eq}), this equation completely determines the general solution (\ref{1/q-expansion}) in the order $(1/q)^1$.

With this in mind, we will make use of the two solutions for $G_{LR}^{(0)} = -G_{RL}^{(0)}$, which correspond to the disconnected and connected parts of $S(T)$. For both regimes the solutions for $G^{(0)}$ determine it on the circle $[0, T]$, and we continue $G^{(0)}$ beyond that in the antiperiodic manner as piecewise-constant functions.  Let us discuss the corresponding large $q$ limits in detail. 

\paragraph{Disconnected part of $S(T)$.} To describe this quantity, we choose the replica-diagonal solution for $G^{(0)}$:
\be
G_{LR}^{(0)} = -G_{RL}^{(0)} = 0\,. \label{G^0-slope}
\ee
Using this and (\ref{G_f}) in (\ref{Sigma^1-eq}) yields
\be \label{Sigma^1-eq-slope}
\begin{aligned}
	\Sigma_{\alpha\alpha}^{(1)}&=-\mJ^2\left(2G_{f}\right)^{q-1}\exp\frac{G_{\alpha\alpha}^{(1)}}{G_{f}}\,; \\
	\Sigma_{LR}^{(1)}&=	\Sigma_{RL}^{(1)} = 0\,.
\end{aligned}
\ee 
We introduce a new variable $g_{\alpha\beta}(t_1, t_2)$:
\bea
g_{\alpha\alpha}(t_1, t_2) &:=& \frac{G_{\alpha\alpha}^{(1)}(t_1, t_2)}{G_{\alpha\alpha}^{(0)}(t_1, t_2)} \,;\\
g_{\alpha\beta}(t_1, t_2) &:=& G_{\alpha\beta}^{(1)}(t_1, t_2)\,, \quad \alpha \neq \beta\,.
\eea
In this case the system of equations (\ref{G^1-eq})-(\ref{Sigma^1-eq}) can be written as 
\bea
&& \partial_{t_1}\partial_{t_2}\left(\sgn(t_1-t_2)g_{\alpha\alpha}(t_1,t_2)\right) = 2\mJ^2 \sgn(t_1-t_2) \e^{g_{\alpha\alpha}(t_1,t_2)}\,; \label{EOM-slope-diag} \\
&& \partial_{t_1}\partial_{t_2}g_{\alpha\beta}(t_1,t_2) = 0,\;\alpha\neq\beta\,. \label{EOM-slope-offdiag}
\eea
This is the final form of saddle point equations for the disconnected spectral form factor. 

\paragraph{Connected part of $S(T)$.} In this case for $G^{(0)}$ we choose the replica-nondiagonal solution\footnote{The exact value of the right hand side constant in (\ref{G^0-ramp}) is chosen from convenience. Note that in \cite{Maldacena18} the same $G^{(0)}_{LR}$ arose as the true two-point function of the two interacting free fermion chains.}:
\be
 G_{LR}^{(0)} = -G_{RL}^{(0)} = \frac{i}{2}\,.\label{G^0-ramp}
\ee
It will be shown that this solution, being supports nondiagonal solutions for the subleading order in $1/q$, which describes the ramp regime. We can write (\ref{Sigma^1-eq}) as follows: 
\be 
\begin{aligned}
	\Sigma_{\alpha\beta}^{(1)}&=s_{\alpha\beta}\mJ^2\left(2G_{\alpha\beta}^{(0)}\right)^{q-1}\exp\frac{G_{\alpha\beta}^{(1)}}{G_{\alpha\beta}^{(0)}}\,. \label{Sigma^1-eq-ramp}
\end{aligned}
\ee 
We use a slightly different definition of $g_{\alpha\beta}(t_1, t_2)$:
\be
g_{\alpha\beta}(t_1, t_2) := \frac{G_{\alpha\beta}^{(1)}(t_1, t_2)}{G_{\alpha\beta}^{(0)}(t_1, t_2)} \,.
\ee
Using this notation and combining (\ref{G^1-eq}) with (\ref{Sigma^1-eq}), we arrive at the final form of the saddle point equation for the connected spectral form factor: 
\be
\partial_{t_1}\partial_{t_2}\left[ G^{(0)}_{\alpha\beta}(t_1,t_2) g_{\alpha\beta}(t_1, t_2)\right] = - s_{\alpha\beta}\mJ^2 \left(2G_{\alpha\beta}^{(0)}\right)^{q-1}\e^{g_{\alpha\beta}(t_1, t_2)}\,. \label{EOM-ramp}
\ee
This equation has the form of the generalized Liouville equation, which is common for SYK computations at large $q$ \cite{MScomments,Choi19,Streicher19,Qi18,Nosaka19,Maldacena18,Jiang19}. Substituting the explicit expressions for $G^{(0)}$, one get the system
\bea \label{EOM-ramp-diag}
&&\partial_{t_1}\partial_{t_2}\left(\sgn(t_1-t_2)g_{\alpha\alpha}(t_1,t_2)\right)=2\mJ^2 \sgn(t_1-t_2) \e^{g_{\alpha\alpha}(t_1,t_2)}; \\
&&\partial_{t_1}\partial_{t_2}g_{\alpha\beta}(t_1,t_2)=2\mJ^2 \e^{g_{\alpha\beta}(t_1,t_2)},\;\alpha\neq\beta. \label{EOM-ramp-offdiag}
\eea

\section{The slope region}
\label{sec:Slope}

\subsection{Large $q$ ansatz}

Let us first focus on the disconnected part of the spectral form factor. The corresponding field configurations, which contribute to the path integral (\ref{S(T)-bilocal}), have uncorrelated $L$ and $R$ copies of the SYK chain, and therefore we want to look for a replica-diagonal solution for $G_{\alpha\beta}$. In the order zero of the $1/q$-expansion we have the solution (\ref{G^0-slope}), and the saddle point equations have the form (\ref{EOM-slope-diag})-(\ref{EOM-slope-offdiag}). Let us now derive the effective action for the field $g_{\alpha\beta}$. The expansion (\ref{1/q-expansion}) for the disconnected spectral form factor can be rewritten as the following ansatz:
\be \label{ansatz-slope}
\begin{aligned}
	G_{\alpha\alpha}(t_1, t_2) &= \frac12 \sgn(t_1 - t_2) \left(1 + \frac{g_{\alpha\alpha} (t_1, t_2)}{q} + o\left(\frac1q\right) \right)\,, \\
	G_{LR}(t_1, t_2) &= \frac{g_{LR} (t_1, t_2)}{q} + o\left(\frac1q\right)\,,\\
	G_{RL}(t_1, t_2) &= \frac{g_{RL} (t_1, t_2)}{q} + o\left(\frac1q\right)\,; \\
	\Sigma_{\alpha\beta}(t_1, t_2) &= \frac{\Sigma^{(1)}_{\alpha\beta}(t_1, t_2)}{q}+ o\left(\frac1q\right)\,.
\end{aligned}
\ee
We substitute these expansions into (\ref{action}) and extract the leading nontrivial action for the fields $g_{\alpha\beta}(t_1, t_2)$ and $\Sigma^{(1)}_{\alpha\beta}(t_1, t_2)$. The Pfaffian term reads 
\be
\begin{aligned}
	T_1 &:= -\frac12 \Tr \log [\delta_{\alpha\beta} \dd_t - \hat{\Sigma}_{\alpha\beta}] =  -\Tr \log (\dd_t) + \frac12 \Tr \frac{1}{q} (\hat{G}_f \cdot \hat{\Sigma}^{(1)}_{\alpha\beta})+\\ &+ \frac12 \Tr \frac{1}{2q^2} (\hat{G}_f \cdot \hat{\Sigma}^{(1)}_{\alpha\gamma} \cdot \hat{G}_f \cdot \hat{\Sigma}^{(1)}_{\gamma\beta}) + O\left(\frac{1}{q^3}\right)\,. 
\end{aligned}
\ee
The polynomial part of the action reads 
\be
\begin{aligned}
	& T_2:= \frac{1}{2}\int_0^T \int_0^T dt_1 dt_2  \left(\Sigma_{\alpha\beta}(t_1, t_2) G_{\alpha\beta}(t_1, t_2) - \frac{2^{q-1} \mJ^2}{q^2}s_{\alpha\beta}  G_{\alpha\beta}(t_1, t_2)^q\right) = \\&= 
	\frac{1}{2}\int_0^T \int_0^T dt_1 dt_2 \frac{1}{q} \Sigma^{(1)}_{\alpha\alpha}(t_1, t_2) G_f(t_1, t_2) + \frac{1}{2}\int_0^T \int_0^T dt_1 dt_2 \frac{1}{q^2} \Sigma^{(1)}_{\alpha\alpha}(t_1, t_2) G_f(t_1, t_2) g_{\alpha\alpha} (t_1, t_2)+ \\
	&+ \frac{1}{2}\sum\limits_{\alpha\neq\beta}\int_0^T \int_0^T dt_1 dt_2 \frac{1}{q^2} \Sigma^{(1)}_{\alpha\beta}(t_1, t_2) g_{\alpha\beta} (t_1, t_2)+\frac12 \int_0^T \int_0^T dt_1 dt_2 \frac{2^{q-1} \mJ^2}{q^2} \frac{1}{2^q}\left(1 + \frac{g_{\alpha\alpha}(t_1, t_2)}{q}\right)^q+ o\left(\frac{1}{q^2}\right)\,.
\end{aligned}
\ee
The $1/q$ terms in $T_1$ and $T_2$ cancel out, and we can carry out the Gaussian integral over $\Sigma^{(1)}$ similarly to the analogous derivations in \cite{Cotler16,Maldacena18,Nosaka19}. the resulting non-vanishing action in the double-scaling limit (\ref{DS}) reads 
\bea
	I_{\text{DS}}[g] &=& \frac{1}{4 \lambda} \int_0^T \int_0^T dt_1 dt_2 \sum_\alpha \bigg(\frac14 \dd_{t_1}\left(\sgn(t_1-t_2) g_{\alpha\alpha}(t_1, t_2)\right) \dd_{t_2}\left(\sgn(t_1-t_2) g_{\alpha\alpha} (t_1, t_2)\right)+\nn\\
	&+& \sum\limits_{\alpha\neq\beta} \dd_{t_1} g_{\alpha\beta}(t_1, t_2) \dd_{t_2} g_{\alpha\beta} (t_1, t_2) + \mJ^2 \e^{g_{\alpha\alpha}(t_1, t_2)}\bigg)\,. \label{Ids-slope}
\eea
The equations of motion for this action are given by (\ref{EOM-slope-diag}),(\ref{EOM-slope-offdiag}). Note the absence of the exponential potential term for the offdiagonal components of $g$. 

\subsection{Early-time solution}
\label{sec:SlopeEarly}

\begin{figure}[t]
\centering 
\includegraphics[scale=0.2]{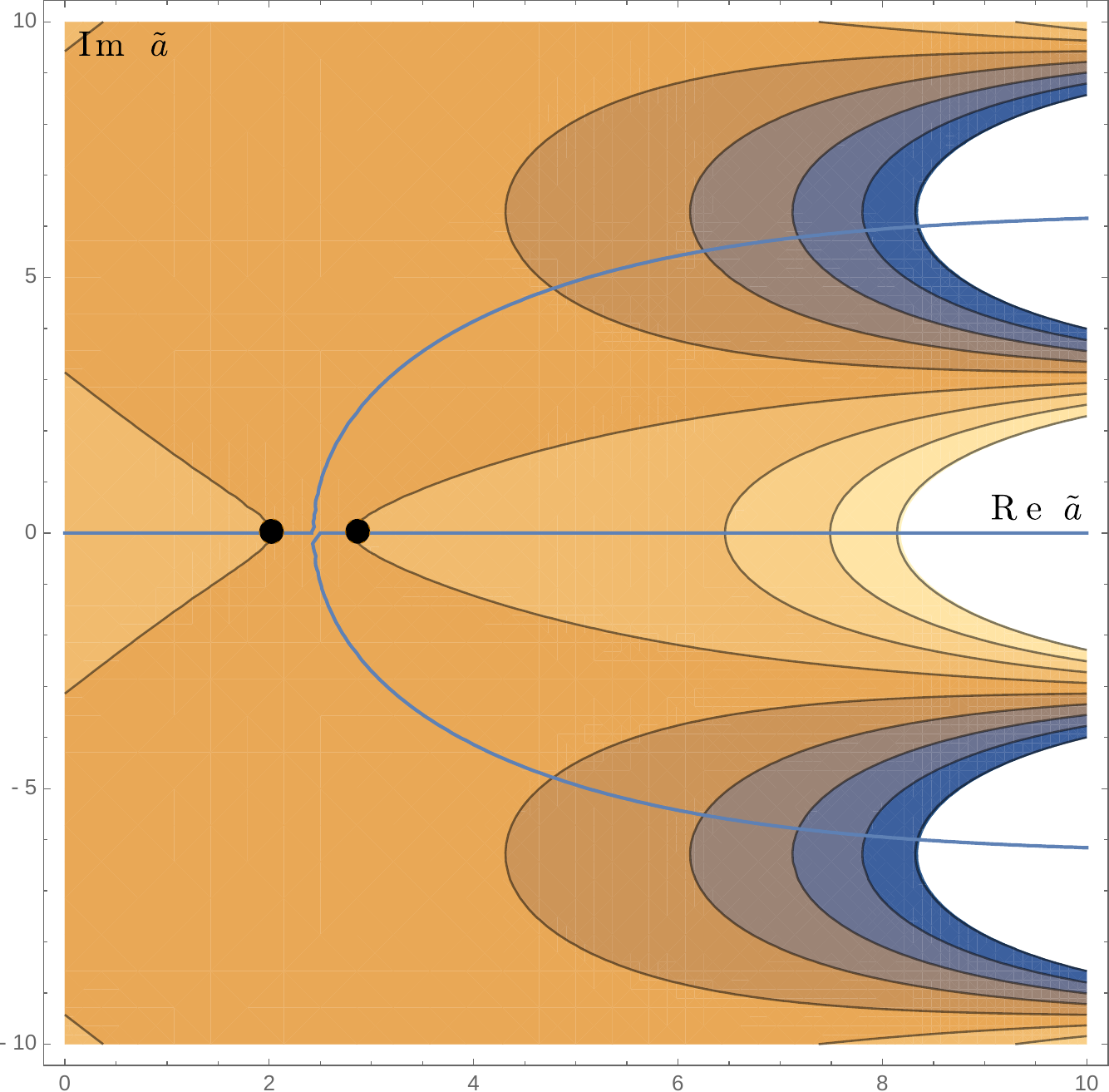}
\includegraphics[scale=0.2]{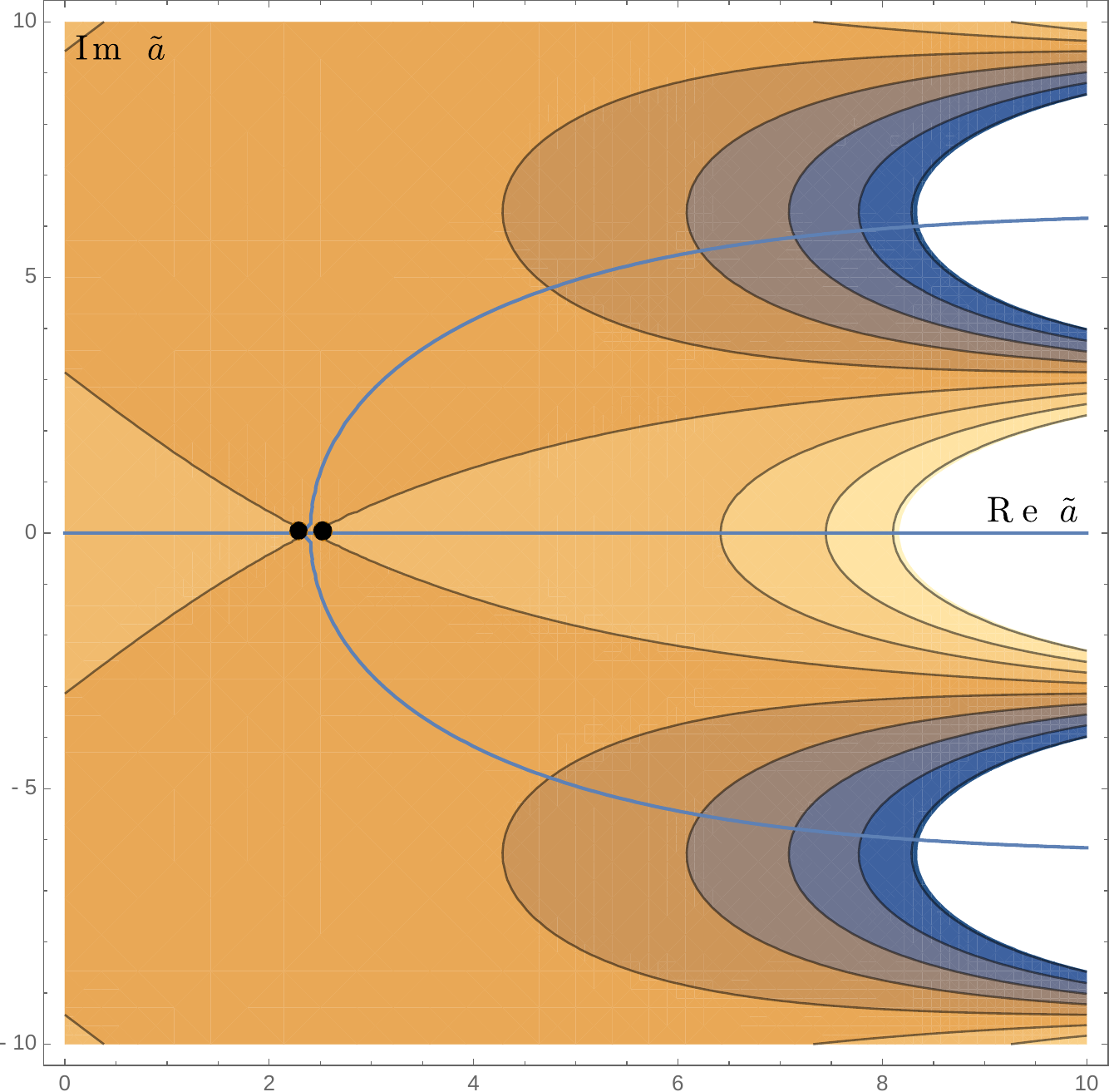}
\includegraphics[scale=0.2]{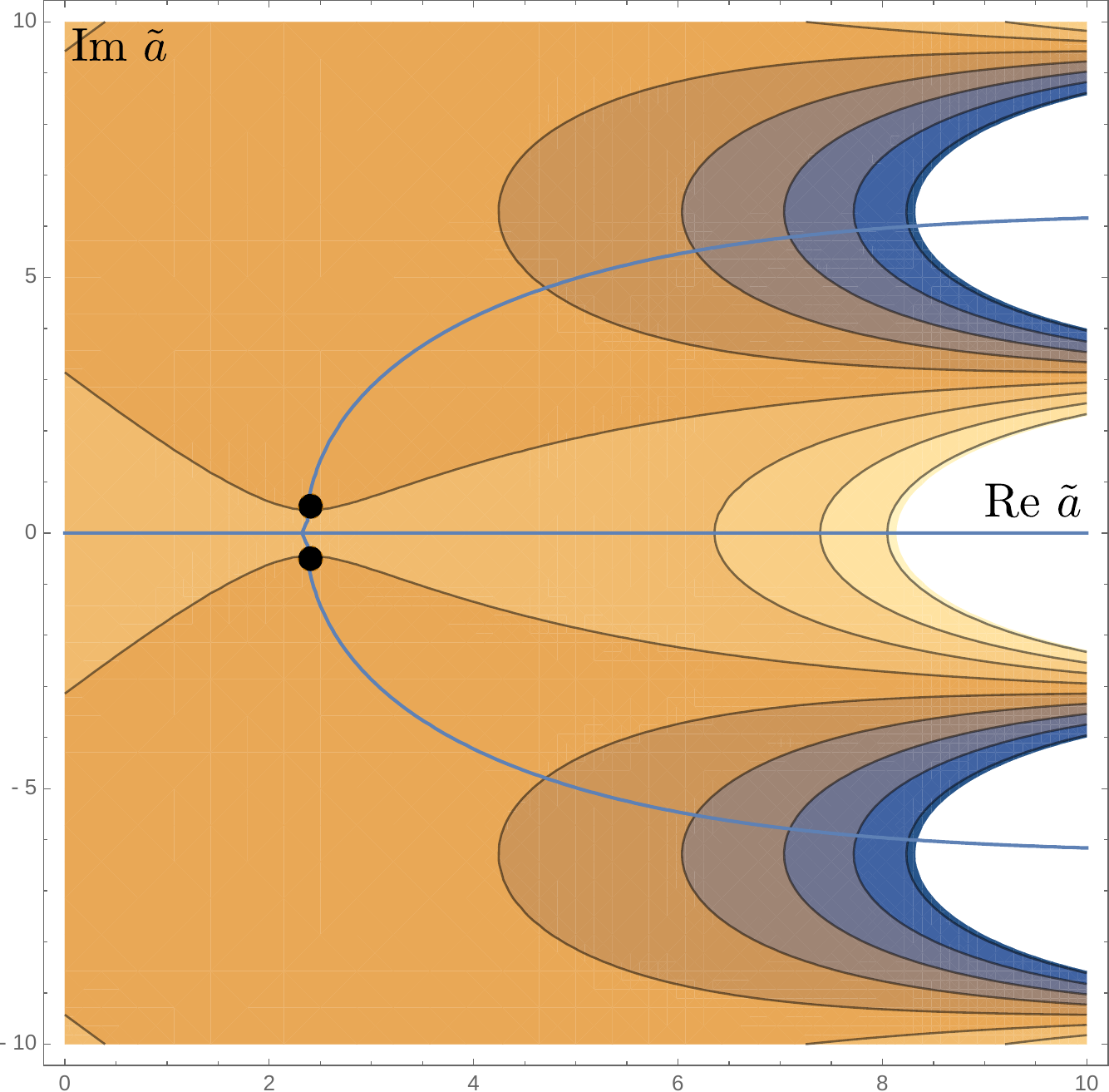}
\includegraphics[scale=0.2]{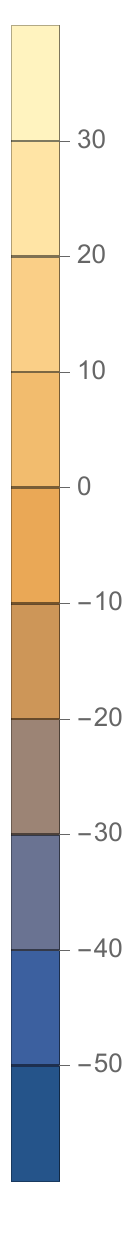}\\
(a) \hspace{4cm} (b) \hspace{4cm} (c)
\caption{Transition from real to complex leading solutions at $T_{\text{cr}}$. Shown are the contour plots of the real part of the constraint (\ref{constr-slope}) together with the contour line given by $\Im$(constraint)$=0$, shown by the blue curves. (a) $\mJ T=1.3$; (b) $\mJ T=1.325$; (c) $\mJ T=1.360$.}
\label{fig:Slope-transition}
\end{figure}

Let us look for solutions with translational symmetry\footnote{Generally speaking, with $\beta \neq 0$ one would expect solutions that are not translationally invariant.}, such that $g_{\alpha\beta}$ depends only on $t=t_1-t_2$. Therefore, the saddle point equations (\ref{EOM-slope-diag})-(\ref{EOM-slope-offdiag}) read
\begin{equation} \label{saddle-slope}
	\begin{aligned}
		&\partial_{t}^2\left(\sgn(t)g_{\alpha\alpha}(t)\right)=-2\mJ^2\sgn(t) \e^{g_{\alpha\alpha}(t)}, \\
		&\partial_{t}^2g_{\alpha\beta}(t)=0,\;\alpha\neq\beta.
	\end{aligned}
\end{equation}
The general solution for the diagonal component reads\footnote{It is the Lorentzian analogue of the general solution for the original large $q$ SYK in Euclidean signature presented in \cite{MScomments}.}:
\be
\e^{g_{\alpha\alpha}(t)} = \frac{a_\alpha^2}{\mJ^2 \cosh^2 (a_\alpha |t| + b_\alpha)}\,, \label{genSol-diag}
\ee
where $a_\alpha$ and $b_{\alpha}$ are integration constants. If $a_L = a_R$, we have a replica-symmetric solution with $g_{LL} = g_{RR}$, otherwise the solution breaks replica symmetry. For the off-diagonal component, the general solution is simply the linear function: 
\be
g_{LR}(t)= -g_{RL}(t) = d + c t\,. \label{genSol-offdiag}
\ee
We see that the general solution is not periodic in $T$, if we assume the same parameters on the entire time circle. This is the key distinction of the large $q$ limit in real time compared to the Euclidean case. Because of this, it is more convenient for us to solve instead on the segment $t \in [0, T/2]$ and then continue the solution to the segment $[T/2, T]$ using the condition
\be
G_{\alpha\beta}(t) = G_{\beta\alpha}(T-t)\,, \label{G-cont}
\ee
which follows from antiperiodicity and antisymmetry of $G_{\alpha\beta}$. Now let us restrict ourselves to the segment $t \in [0, T/2]$ and impose the boundary conditions on its endpoints for $g$. The first boundary condition is determined by the free fermions being the UV limit of SYK:
\be \label{conds-slope-1}
g_{\alpha\alpha}(0)=0\,.
\ee
The second boundary conditions follows from (\ref{G-cont}) as well as from smoothness at $t=T/2$: 
\bea \label{conds-slope-2}
g_{\alpha\alpha}'\left(\frac{T}{2}\right)=0 \,; \qquad
g_{LR}\left(\frac{T}{2}\right)=0\,.
\eea
Taking into account the boundary conditions (\ref{conds-slope-1}),(\ref{conds-slope-2}), we arrive at the short time solution for $t \in \left[0, \frac{T}{2}\right]$ in the following form:
\bea\label{slope-sol-diag}
	\mathrm{e}^{g_{\alpha\alpha}(t)}&=&\left\{\frac{\cosh \frac{\tilde{a}_{\alpha}}{2}}{\cosh \left[\tilde{a}_{\alpha}\left(\frac{1}{2}-\frac{t}{T}\right)\right]}\right\}^{2}\,;\\\label{slope-sol-offdiag}
	g_{LR}(t) &=& - g_{RL}(t) = c\left(t - \frac{T}{2}\right)\,.
\eea
Here the parameter $\tilde{a}_{\alpha}:=a_{\alpha} T$ has to solve an algebraic constraint
\be
	\quad \tilde{a}_{\alpha}^{2}=(\mathcal{J} T)^{2} \cosh ^{2} \frac{\tilde{a}_{\alpha}}{2}\,. \label{constr-slope-sq}
\ee
Before proceeding further, let us note that the solution (\ref{slope-sol-diag}) has a $\ZZ_2$-symmetry in each SYK copy: it is invariant under replacement $\tilde{a}_\alpha$ $\to$ $-\tilde{a}_\alpha$. The initial condition as written in the form of (\ref{constr-slope-sq}) preserves this symmetry. All further saddle point solutions which we will discuss also will have a similar symmetry. Thus here and henceforth we can essentially mod it out from the spectral form factor by considering a square root with the plus sign of the equation (\ref{constr-slope-sq}) as the constraint for the parameter $\tilde{a}$ instead: 
\be
\quad \tilde{a}_{\alpha}=\mathcal{J} T \cosh \frac{\tilde{a}_{\alpha}}{2}\,; \qquad \Re\ \tilde{a}_\alpha > 0\,. \label{constr-slope}
\ee
Let us analyze this constraint in more detail. Unlike the Euclidean case \cite{MScomments}, it does not have a real solution for all values of $\mJ T$. Instead, there is a critical time $T_{\text{cr}} \simeq 1.33/\mJ $ which separates two regimes: 
\begin{itemize}
	\item For $T < T_{\text{cr}}$, there are two real positive solutions $\tilde{a}_1$ and $\tilde{a}_2$, with $\tilde{a}_1 < \tilde{a}_2$, as shown on Fig. \ref{fig:Slope-transition}(a). At $T = T_{\text{cr}}$ they coalesce into a unique solution, as shown on Fig. \ref{fig:Slope-transition}(b). Besides these two, there is an infinite family of complex solutions, for which $\Re\ \tilde{a} > \tilde{a}_2$.
	\item For $T > T_{\text{cr}}$, there are now two new complex-valued solutions, $\tilde{a}_1$ and $\tilde{a}_2 = \tilde{a}_1^*$, as shown on Fig. \ref{fig:Slope-transition}(c). Other complex solutions have $\Re\ \tilde{a} > \Re\ \tilde{a}_1$.
\end{itemize}

As we will see in section \ref{sec:Slope-action}, the solutions with the lowest $\Re\ \tilde{a}$, the ones experiencing the transition from real to complex, actually correspond to the leading saddle points of the spectral form factor path integral at all times, and other complex solutions correspond to subleading saddle points. Correspondingly, we will refer to solutions with the lowest $\Re\ \tilde{a}$ as leading solutions, and to other solutions with higher $\Re\ \tilde{a}$ as subleading ones. 

\begin{figure}[t]
	\centering 
	\includegraphics[scale=0.15]{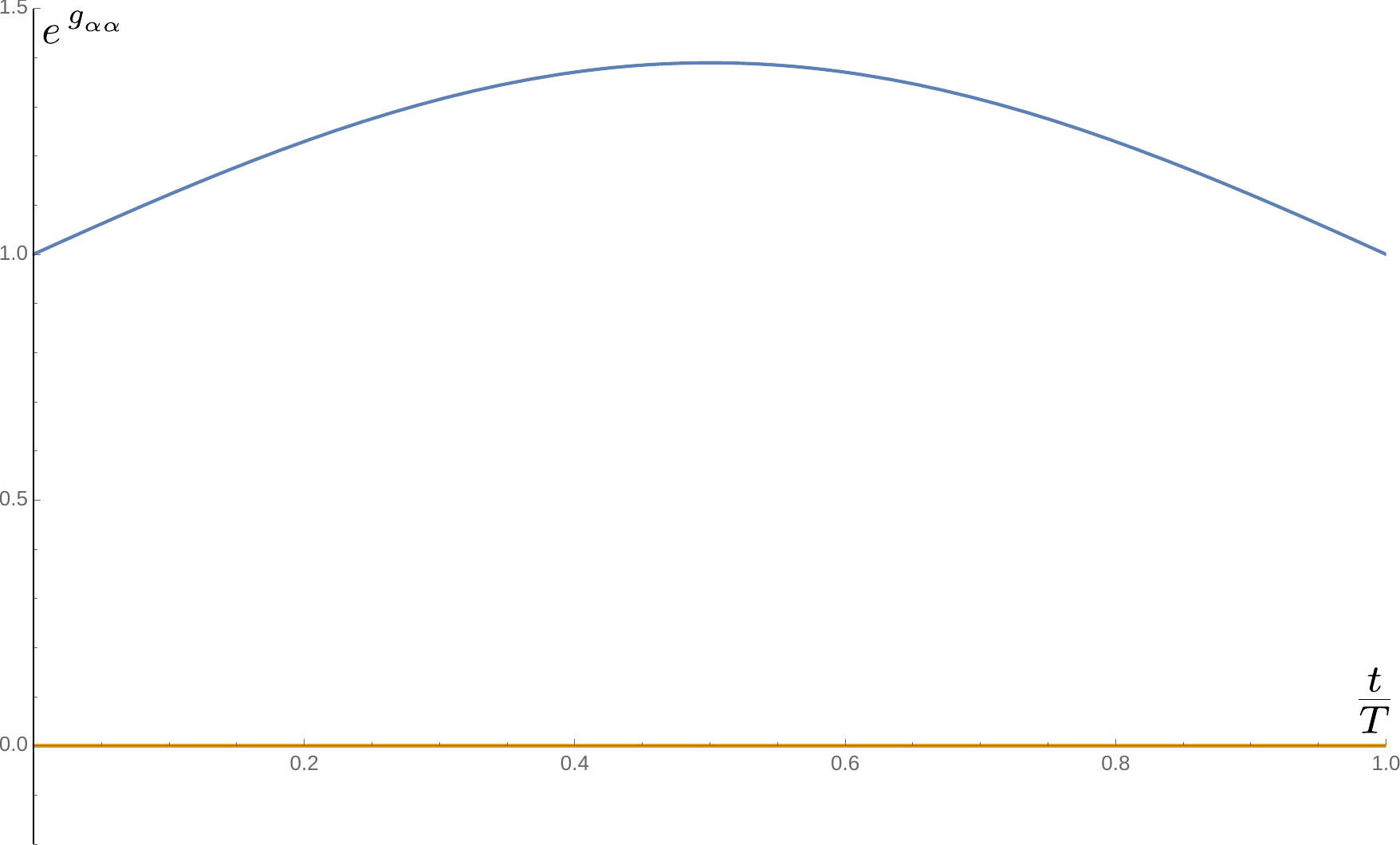}$\quad$
	\includegraphics[scale=0.15]{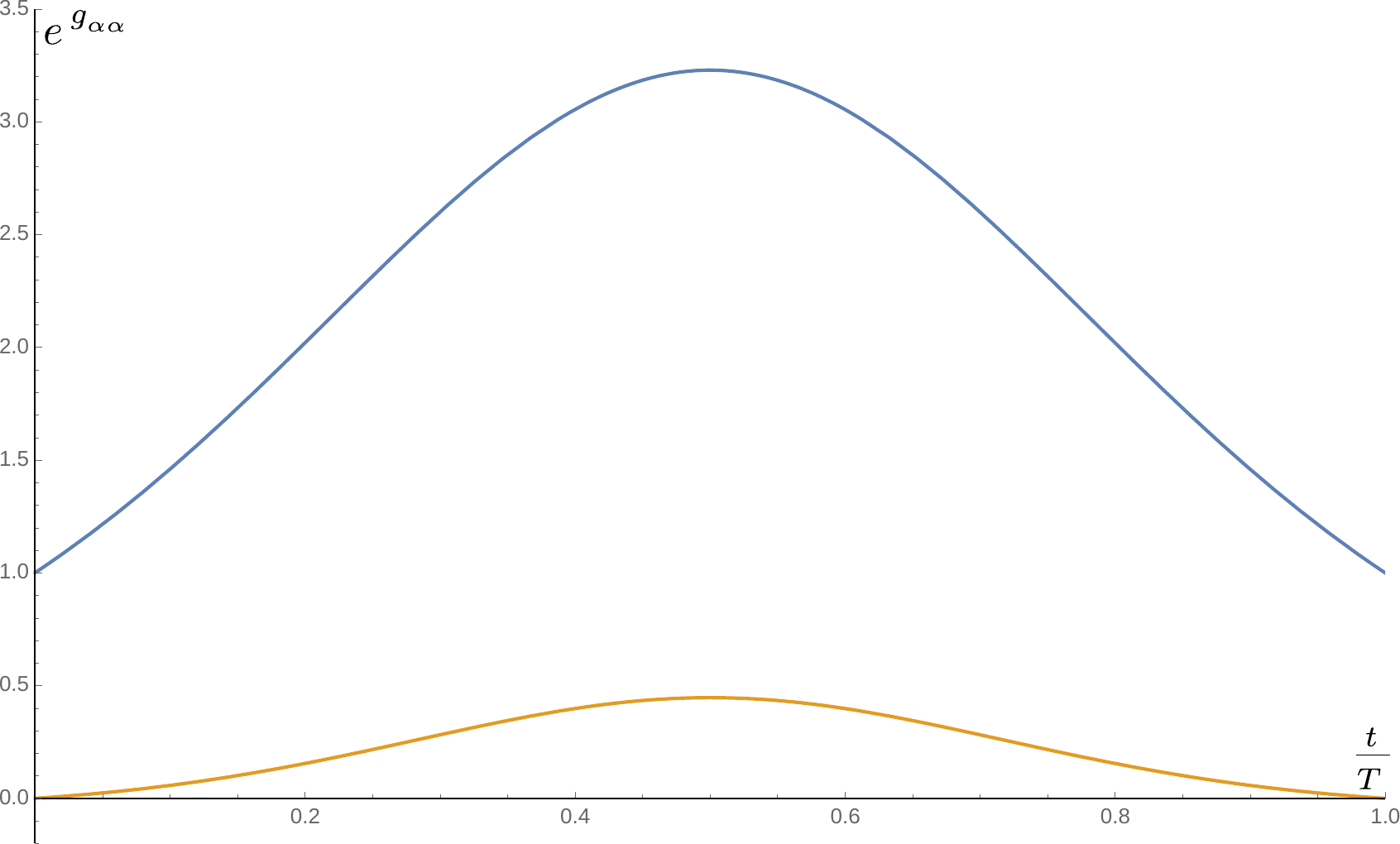}$\quad$
	\includegraphics[scale=0.15]{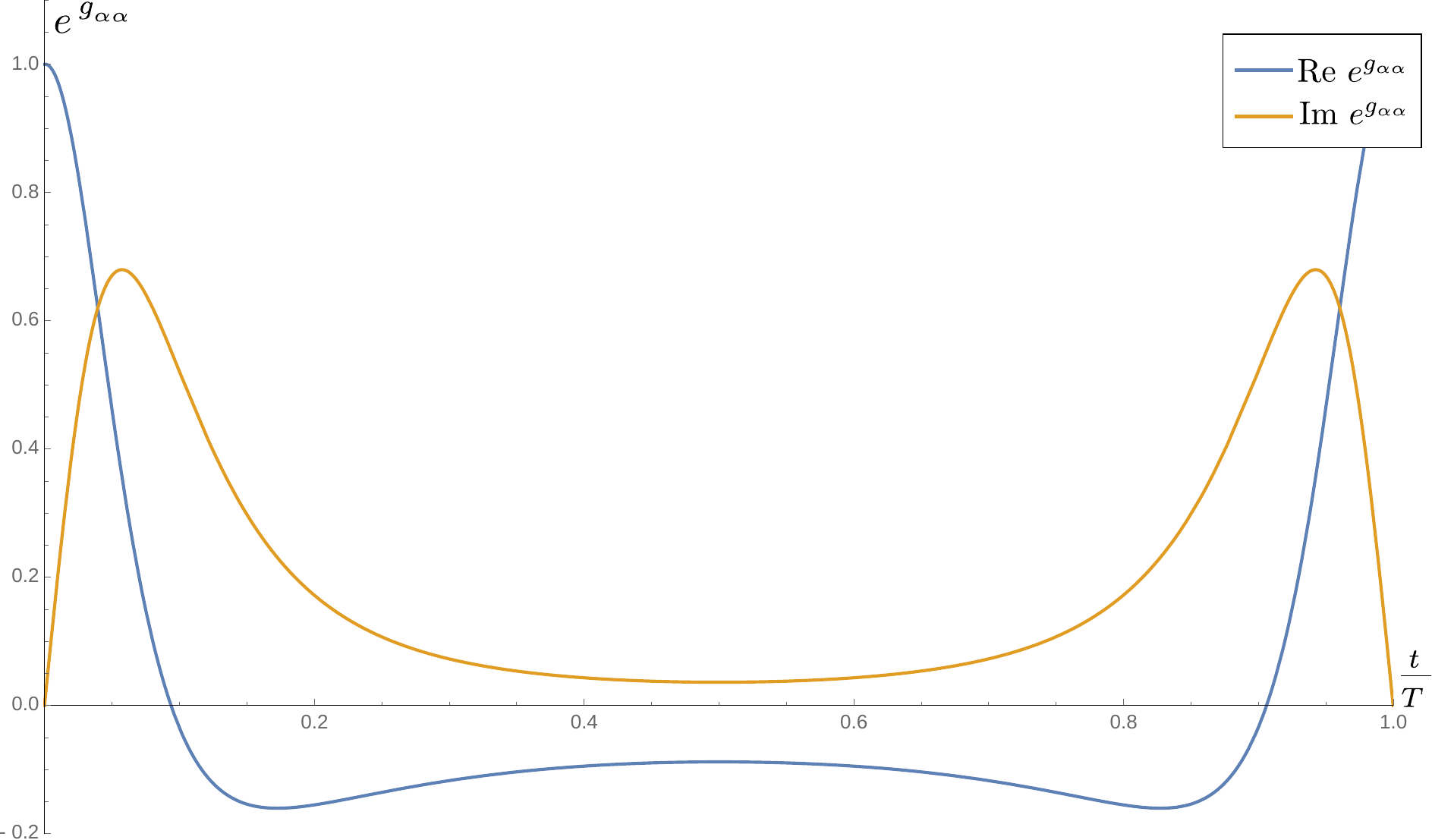}$\quad$\\
	(a) \hspace{4cm} (b) \hspace{4cm} (c)\\
	\includegraphics[scale=0.14]{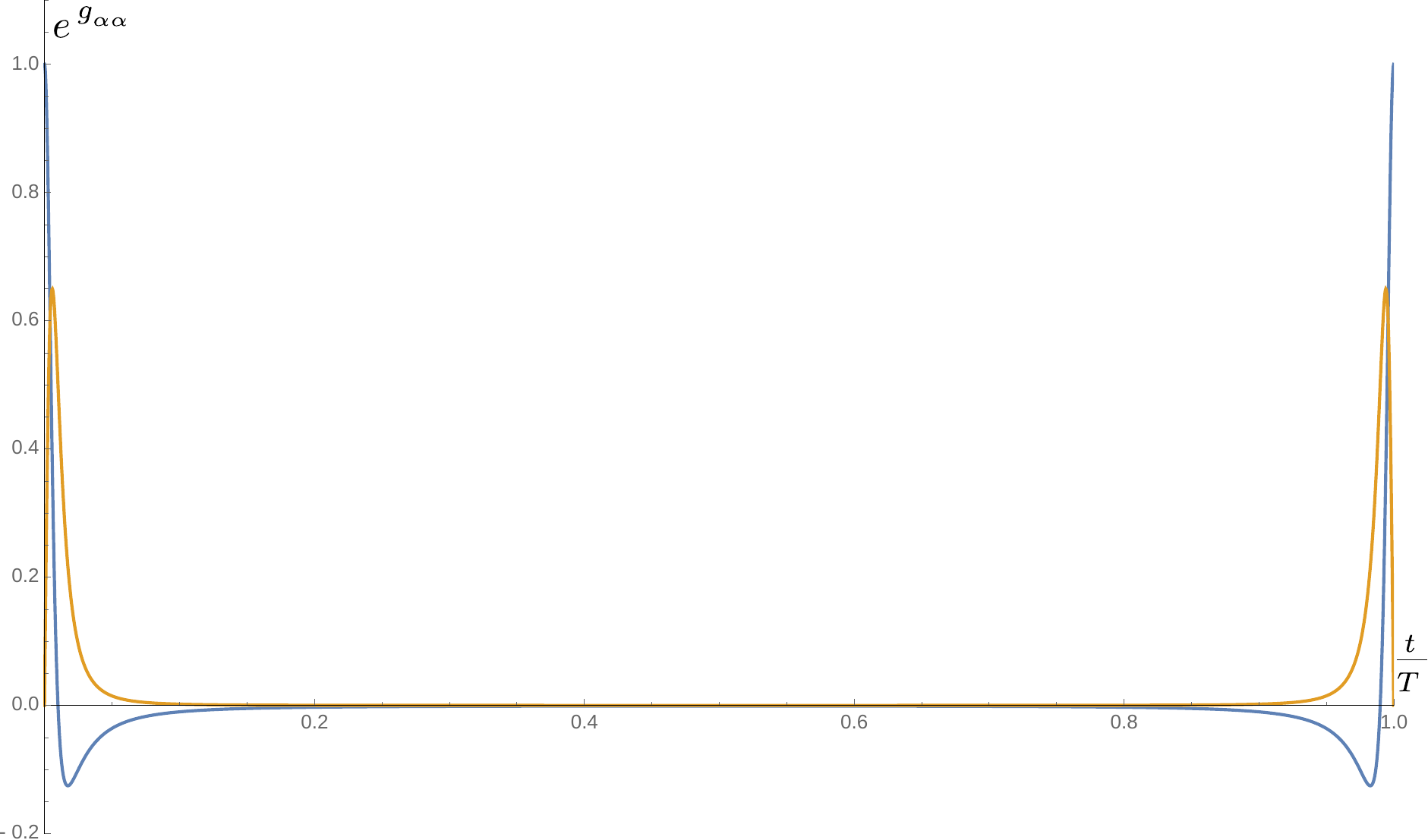}$\quad$
	\includegraphics[scale=0.14]{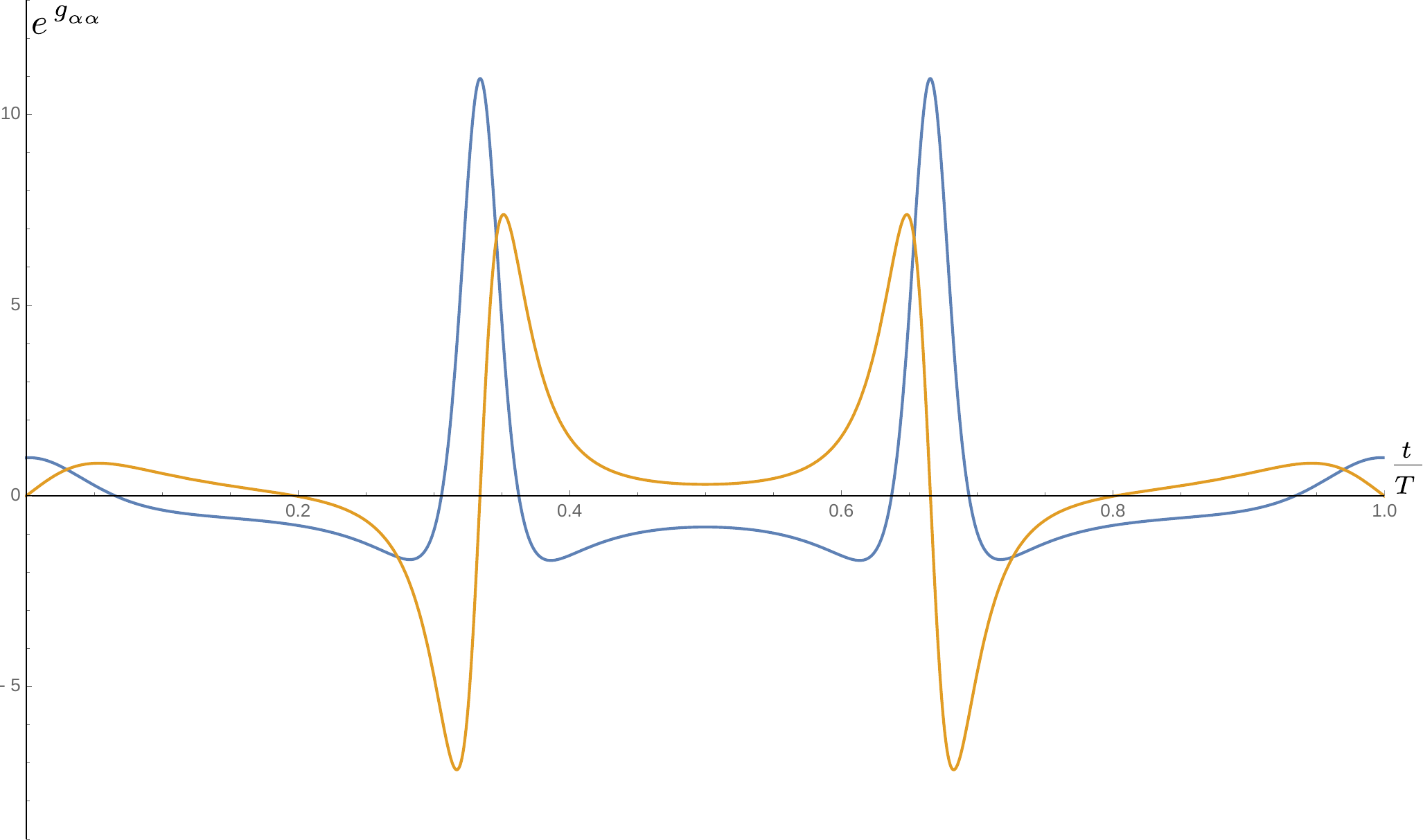}$\quad$
	\includegraphics[scale=0.13]{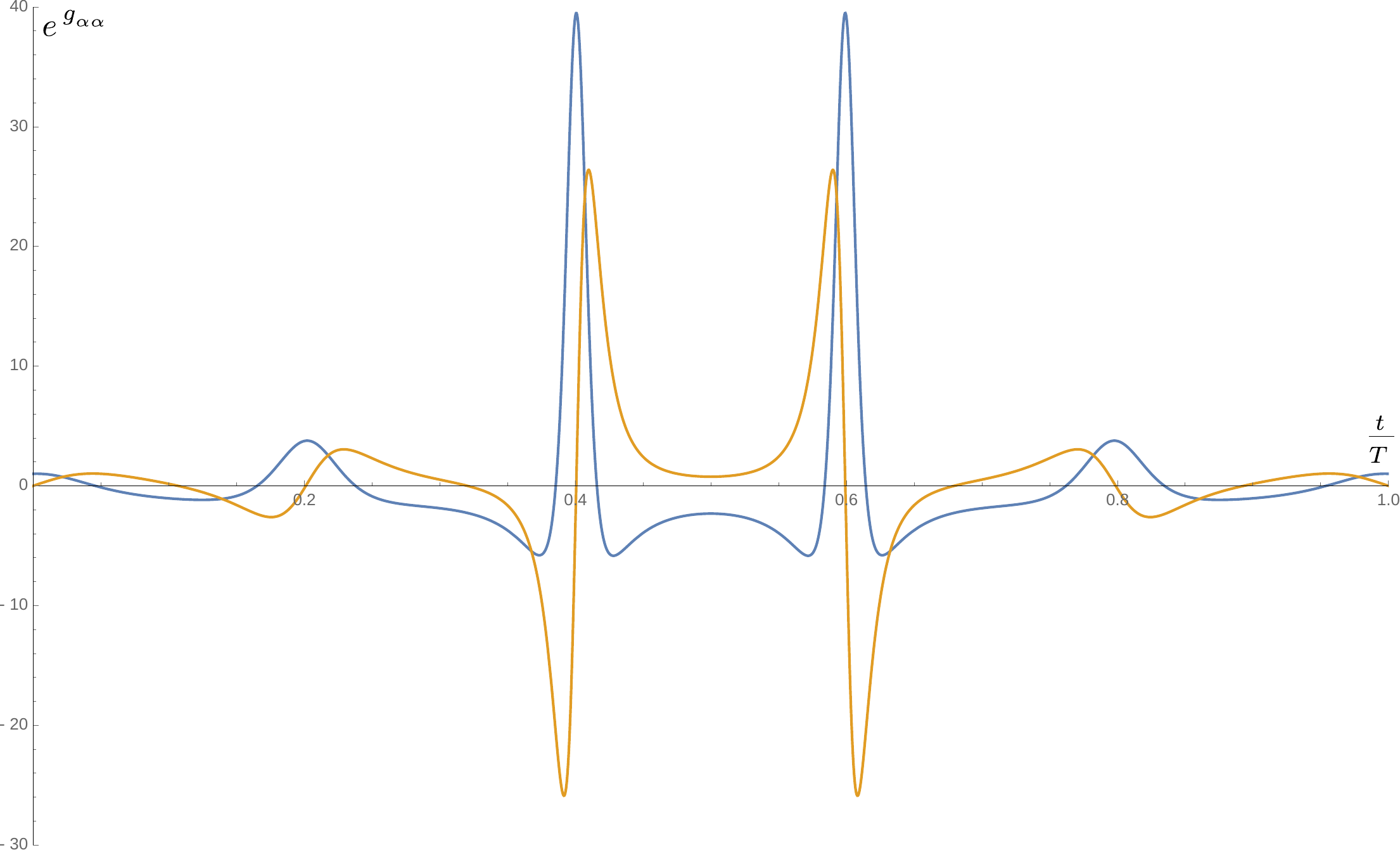}\\
	(d) \hspace{4cm} (e) \hspace{4cm} (f)
\caption{The solution (\ref{slope-sol-diag}) for different times and different solutions of the constraint (\ref{constr-slope}). (a) $\mJ T=1$, $\tilde{a}\simeq 1.18$; (b) $\mJ T=1.33$, $\tilde{a}\simeq 2.4+0.17 i$; (c) $\mJ T=10$, $\tilde{a}\simeq 0.60+3.03 i$; (d) $\mJ T=100$, $\tilde{a}\simeq 0.06+3.14 i$; (e) $\mJ T=10$, $\tilde{a}\simeq 1.65+9.18 i$; (f) $\mJ T=10$, $\tilde{a}\simeq 2.45+15.44 i$}
\label{fig:SlopeSolutions}
\end{figure}

We demonstrate the behavior of $\e^{g_{\alpha\alpha}}$ as determined by the solution (\ref{slope-sol-diag}) for values of $\tilde{a}$ which satisfy the constraint on Fig. \ref{fig:SlopeSolutions} for different values of $\mJ T$. The plots (a)-(d) show leading solutions, and the plots (e),(f) show subleading solutions. For $T > T_{\text{cr}}$ the solution acquires non-zero imaginary part (shown by the orange curve), as can be seen on the plots (b)-(d). The subleading solutions with higher $\Re\ \tilde{a}$ acquire extra extrema compared to the leading solutions, as shown on the plots (e) and (f). This is similar to subleading saddle points in the Euclidean SYK partition function at finite $q$ \cite{Cotler16,AKTV,AKV}. 

Since we are primarily interested in the late time behavior, let us solve the constraint (\ref{constr-slope}) for late times analytically (assuming the validity of the $1/q$-expansion).  Denote $x=\Re(\tilde{a})$ and $y=\Im(\tilde{a})$, then we have the system of equations:
\begin{equation}\label{system}
	\begin{cases}
		\cos\left(\frac{y}{2}\right)\cosh\left(\frac{x}{2}\right)=\frac{x}{\mJ T}\,; \\
		\sin\left(\frac{y}{2}\right)\sinh\left(\frac{x}{2}\right)=\frac{y}{\mJ T}\,.
	\end{cases}
\end{equation}
Let us look for a solution as expansion in $(\mJ T)^{-1}$. In the first order we can set the right hand parts of the above equations to zero and get:
\begin{equation}
	\begin{cases}
		y=\pm (2n+1 )\pi,\quad n\in \mathbb{Z}_+\,; \\
		\frac{x}{2}=\frac{|y|}{\mJ T}=\frac{(2n+1)\pi}{\mJ T}\,.
	\end{cases}
\end{equation}
Now let us find the correction to $y$. Putting $y=\pm[(2n+1)\pi -\epsilon]$ into the first equation in~\eqref{system}, we get:
\begin{equation}
	\frac{\epsilon}{2}=\frac{x}{\mJ T}=2\frac{(2n+1) \pi}{(\mJ T)^2}.
\end{equation}
So we can write down the solution as
\begin{equation}\label{bigJT-a}
	\tilde{a}=2\frac{(2n+1) \pi}{\mJ T}\pm\left((2n+1) \pi-4\frac{(2n+1) \pi}{(\mJ T)^2}\right)i+o\left(\frac{1}{(\mJ T)^2}\right)\,,\quad n \in \ZZ_+\,.
\end{equation}
\begin{figure}[t]
	\center{\includegraphics[scale=0.25]{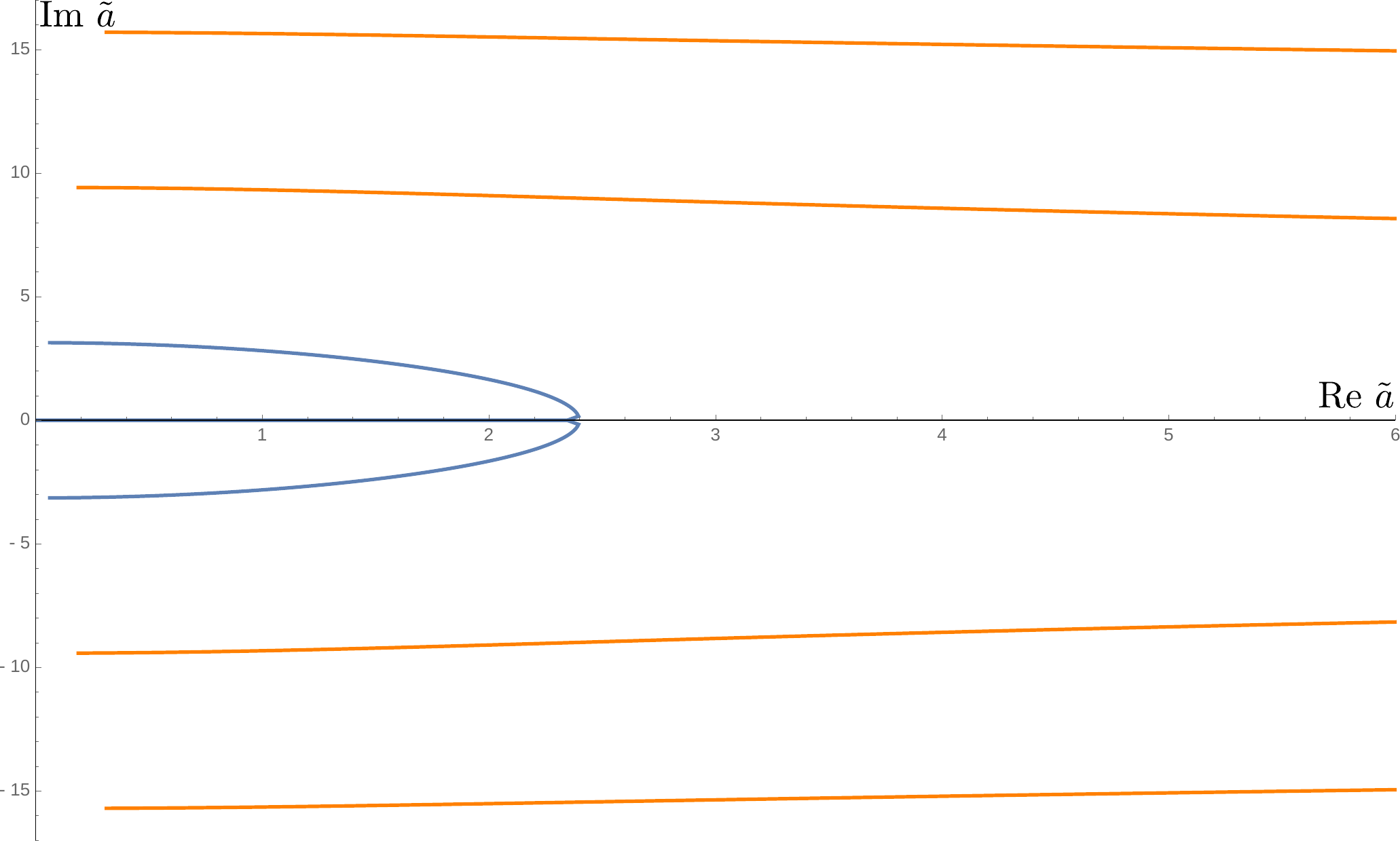}}\caption{Locations of solutions of the constraint (\ref{constr-slope}) on the complex right half-plane parametrized by $\mJ T \in [0, 100]$. The trajectories of subleading solutions (shown by orange curves) are not shown completely, but only their late-time tail ends are shown.}\label{fig:Slope-trajectories}
\end{figure}
The case $n = 0$ corresponds to the leading solution, whereas $n > 0$ correspond to subleading solutions. It is worth noting that at late times $\Re\ \tilde{a} \to 0$, whereas $\Im\ \tilde{a} \to \pm i \pi(2n+1)$. We show the dependence of solutions of the constraint (\ref{constr-slope}) on general $\mJ T$ on the Fig. \ref{fig:Slope-trajectories}. The leading solutions are shown by the blue curve, and the orange curves represent some of the subleading solutions. 

The question we need to answer next is what kind of contributions do these solutions introduce to $S(T)$. Before we do that, let us remember that the entire above discussion is based upon the $1/q$-expansions (\ref{1/q-expansion}), which might break down at late times. So let us address the region of late times.

\subsection{Late-time solution}
\label{sec:SlopeLate}

To study the solutions of saddle point equations \eqref{EOM} at late times beyond the expansion (\ref{1/q-expansion}), we will use the approach similar to one used in the studies of the SYK dual to the traversable wormhole \cite{Maldacena18} and similar models \cite{Nosaka19}. The key observation is the following. From the second equation in~\eqref{EOM} we see that $\Sigma_{\alpha\beta}$ varies $q$ times faster than $G_{\alpha\beta}$, but in this case $\Sigma_{LR}=\Sigma_{RL}=0$ and $\Sigma_{\alpha\alpha}$ are odd functions. Therefore, at very long times, we can approximate $\Sigma_{\alpha\alpha}$ as a derivative of a delta function: $\Sigma_{LL}=\kappa_L \delta'(t)$ and $\Sigma_{RR}=\kappa_R \delta'(t)$. So the equations~\eqref{EOM} can be approximated as
\begin{equation}
	(1-\kappa_{L,R})\partial_t G_{\alpha\beta}=0\,. \label{EOM-slope-late}
\end{equation}
The solutions are constants: 
\be
G_{\alpha\alpha}(t) = A_\alpha\,; \qquad  G_{LR}(t) = -G_{RL}(t) = C\,. \label{slope-lateT}
\ee
Now we have to glue this smoothly to the early-time result which has the form (\ref{ansatz-slope}) with $g_{\alpha\beta}$ given by solutions (\ref{slope-sol-diag}),(\ref{slope-sol-offdiag}). The diagonal components have the form\footnote{This is for large positive times, so $\sgn(t) = 1$.}:
\be\label{largeT-diag}
		G_{\alpha\alpha}(t)=\frac{1}{2}\left(1+\frac{g_{\alpha\alpha}(t)}{q}\right)\,. 
\ee
One can check that for all values of $\tilde{a}$ allowed by the constraint (\ref{constr-slope}) the function $g_{\alpha\alpha}$ stays finite at late times, so to the leading order in $1/q$ the expansion is consistent with the constant late-time solution (\ref{slope-lateT}) with $A_L = A_R = \frac12$. This essentially means that all of the dynamics is captured by the early times, and the corresponding solution (\ref{slope-sol-diag}) is valid for arbitrarily large times. 

Meanwhile, the off-diagonal component is determined by the expansion 
\be
		G_{LR}(t)=-G_{RL}(t)=\frac{g_{LR}(t)}{q}=0\,,
\ee
with the solution (\ref{slope-sol-offdiag}) for $g$. We see that the early-time linear solution can only be consistent with the late-time constant behavior only with  $c = 0$, and, thus $g_{LR}=0$. This is expected as non-diagonal components of $G$ must trivialize for the disconnected part of the spectral form factor. Thus, from the late-time analysis we come to two points: 
\begin{itemize}
	\item The $1/q$-expansion for the disconnected part of the spectral form factor is valid at all times.
	\item The ansatz (\ref{ansatz-slope}) on the saddle point equations yields $G_{LR} = G_{RL} = 0$. 
\end{itemize}

\subsection{On-shell action}
\label{sec:Slope-action}
Since we've shown that the $1/q$-expansion describes the disconnected part of the spectral form factor at all times, we can use the action (\ref{Ids-slope}) for all times. Using the saddle point equations (\ref{EOM-slope-diag}),(\ref{EOM-slope-offdiag}) and the symmetry properties of the solutions, we can rewrite the action as follows:
\be
I_{\text{DS}}[g] = \frac{T\mJ^2}{\lambda} \int\limits_{0}^{T/2} dt\, \e^{g_{\alpha\alpha}(t)}\left(1-\frac12g_{\alpha\alpha}(t) \right)\,.
\ee
Substituting the solution (\ref{slope-sol-diag}) and computing the integral, we get the result
\begin{equation}\label{slope-action}
		\lambda I_{\text{on-shell}} =\sum_{\alpha=L,R} \left(2 \tilde{a}_{\alpha} \tanh \left(\frac{\tilde{a}_{\alpha}}{2}\right)-\frac{\tilde{a}_{\alpha}^{2}}{2}\right)\,.
\end{equation}
where $\tilde{a}_{\alpha}$ solve the constraint (\ref{constr-slope}) independently. This action is complex-valued on complex saddle points. 

We can analyze the action for $\mJ T\rightarrow\infty$ using the solutions for $\tilde{a}_{\alpha}$ from~\eqref{bigJT-a}. Substituting them to~\eqref{slope-action}, we get:
\bea
		&&\Re\, \lambda I_{\text{on-shell}}=\frac{\pi^2}{2}\left((2n+1)^2+(2m+1)^2\right)+O\left(\frac{1}{(\mJ T)^2}\right),\quad n,m\in\mathbb{Z}_+\,; \label{X}\\
		&&\Im\, \lambda I_{\text{on-shell}}=\pm \mJ T\pm \mJ T+O\left(\frac{1}{\mJ T}\right),  \label{Y}   
\eea
where one can choose any combination of signs. We see that the leading contribution to the spectral form factor comes from the solution with $n=m=0$, which is what we referred to as the leading solution, for both $L$ and $R$ copies. The solutions with non-zero $n$ and/or $m$ have larger action and are thus indeed subleading. 
\begin{figure}[t]
	\center{\includegraphics[scale=0.32]{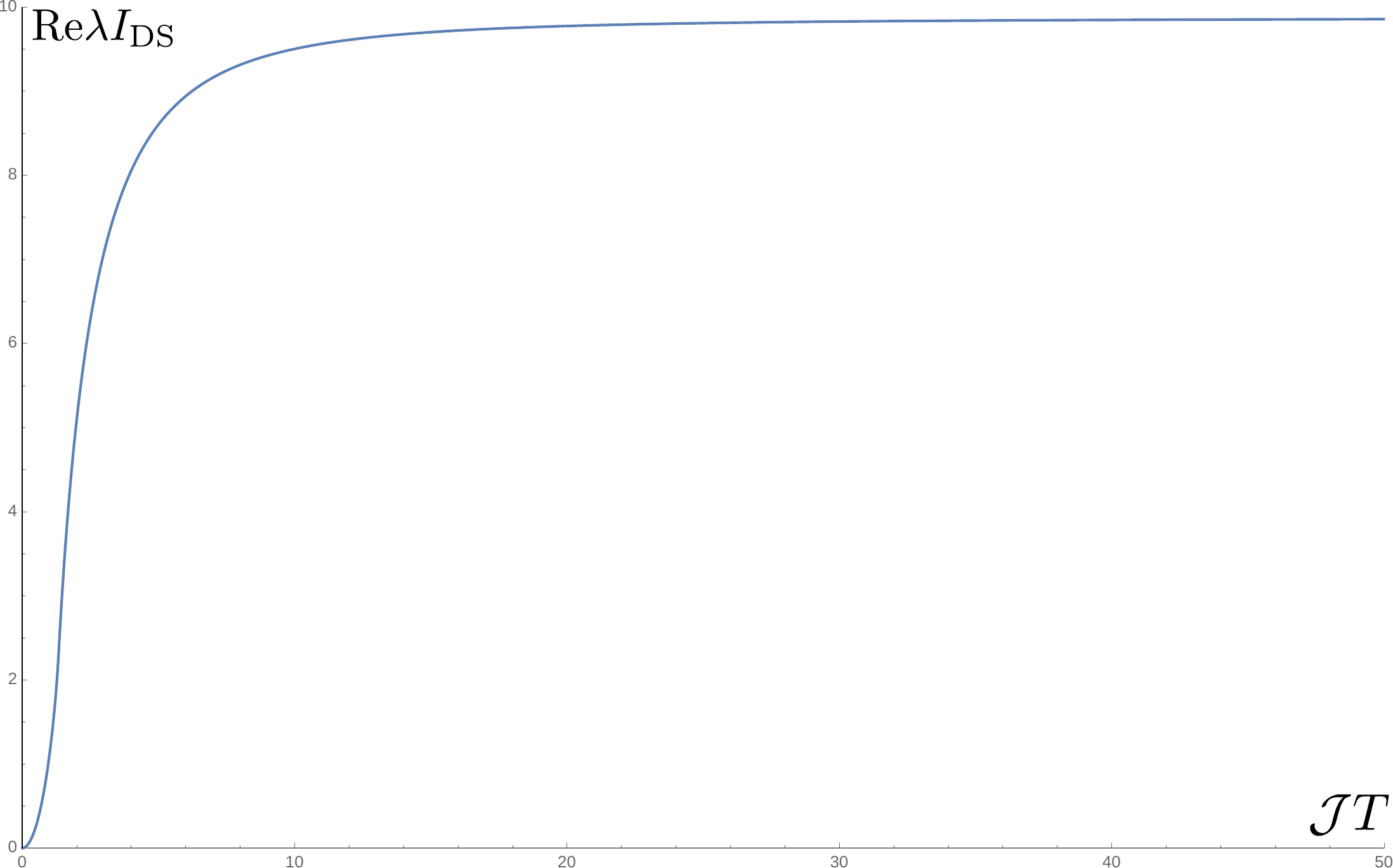}}\caption{Real part of the action on a slope saddle point as a function of time.}
	\label{fig:ActionPlot}
\end{figure}
An important point here is the arrangement of leading solutions between the copies (replicas). Taking the replica-symmetric solution, meaning $\tilde{a}_L = \tilde{a}_R = \tilde{a}_1$, results in the action that has non-zero imaginary part, because the imaginary parts from individual replicas add up instead of canceling each other. Conversely, a replica symmetry breaking solution with $\tilde{a}_L = \tilde{a}_1$ and $\tilde{a}_R = \tilde{a}_1^*$ will give total action that is real-valued. Because of this interpretation, we will sometimes refer to $T_{\text{cr}}$ as replica symmetry breaking time. 

 Another point is that in the region of complex solutions there is a $\ZZ_2 \times \ZZ_2$-symmetry that acts like $\Im\ \tilde{a}_\alpha \to - \Im\ \tilde{a}_\alpha $. The individual solutions (\ref{slope-sol-diag}) and the replica-symmetric saddle points spontaneously completely break this symmetry, while the leading replica symmetry breaking saddles preserve a $\ZZ_2$ subgroup. Hence we can divide the entire contribution from complex solutions by $4$ to mod out this symmetry. 

With these points in mind, we can write down the contribution to the path integral from at the saddle point level for $T > T_{\text{cr}}$. Let us denote as $I[\tilde{a}_L, \tilde{a}_R]$ the on-shell action for a generic solution. Then we see that
\bea
X[\tilde{a}_1]:= \Re\ I[\tilde{a}_1, \tilde{a}_1^*] &=& \Re\ I[\tilde{a}_1^*, \tilde{a}_1] = \Re\ I[\tilde{a}_1, \tilde{a}_1]  = \Re\ I[\tilde{a}_1^*, \tilde{a}_1^*] \,;\\
\Im\ I[\tilde{a}_1, \tilde{a}_1^*] &=& \Im\ I[\tilde{a}_1^*, \tilde{a}_1] = 0\,;\\
Y[\tilde{a}_1]:=\Im\ I[\tilde{a}_1, \tilde{a}_1] &=& -\Im\ I[\tilde{a}_1^*, \tilde{a}_1^*] \,.
\eea
Thus we see that the pair of leading complex-conjugated solutions parametrized by $\tilde{a}_1$ will give the contribution to the spectral form factor equal to 
\bea
S(T) &\to& \frac14 \left[\e^{- I[\tilde{a}_1, \tilde{a}_1^*]}+\e^{- I[\tilde{a}_1, \tilde{a}_1^*]}+\e^{- I[\tilde{a}_1, \tilde{a}_1]}+\e^{- I[\tilde{a}_1^*, \tilde{a}_1^*]}\right] = \e^{-X[\tilde{a}_1]}\frac{1 + \cos Y[\tilde{a}_1]}{2}\nn\\ &=& \e^{-X[\tilde{a}_1]} \cos^2 \frac{Y[\tilde{a}_1]}{2}\,.
\eea
At late times with $X$ and $Y$ given by (\ref{X}) and (\ref{Y}) correspondingly with $n=m=0$, we get the semiclassical result: 
\begin{equation}
\frac{|\langle Z(iT) \rangle|^2}{2^N} \simeq \e^{-\frac{\pi^2}{\lambda}}\cos^2\frac{\mJ T}{\lambda}\,. \label{slope-semiclassical}
\end{equation}
The squared cosine prefactor is similar to the behavior appearing in RMT (specifically, GUE) in the same time region \cite{Gharibyan18}. Here we have obtained it from the saddle point structure of the SYK model in the large $q$ limit, rather than from the spectral density. 

The time dependence of the real part of the action on a leading solution is presented on the Fig. \ref{fig:ActionPlot}. We see that the action increases up to a constant, with the full agreement with expectations from the slope region. However, the actual decay rate is dominated by the quantum corrections \cite{Cotler16,Garcia-Garcia16}, so as a next step we turn the computation of the one-loop quantum correction to the slope saddles we've discussed.

\subsection{One-loop correction}
\label{sec:1loop-slope}

Let us finally calculate spectral form factor on the slope taking into account quantum corrections. We have shown that the disconnected part of the spectral form factor is described by the $1/q$-expansion at all times at the semiclassical level. Hence we can consider the quantum dynamics using the path integral over $g_{\alpha\beta}$ with the action (\ref{Ids-slope}). We restrict ourselves to the one-loop correction in terms of the expansion in $\lambda$. We split the field $g_{\alpha\beta}$ into classical and quantum parts:
\begin{equation}
	g_{\alpha\beta}=g_{\alpha\beta}^{cl}+\mathfrak{g}_{\alpha\beta},
\end{equation}
where $g_{\alpha\beta}^{cl}$ solve the saddle point equations~\eqref{EOM-slope-diag}-(\ref{EOM-slope-offdiag}). We will temporarily use the dimensionless time variables $u =\mJ (t_1-t_2)$ and $v = \mJ \frac{t_1+t_2}{2}$. Expanding the path integral around a single saddle point, we get:
\bea\label{SFFslope}
		&& S(T) \to \e^{-I_{\text{on-shell}}}\int\mathcal{D}\mathfrak{g}_{\alpha\beta} \exp\left\{\frac{1}{4 \lambda} \int\limits_{-\mJ T}^{\mJ T} d u \int\limits_{0}^{\mJ T} d v \sum\limits_{\alpha\neq\beta}\mathfrak{g}_{\alpha\beta}(u,v)\left(\frac{1}{4}\partial^2_{v}-\partial^2_u\right)\mathfrak{g}_{\alpha\beta}(u, v)\right\}\times\\&&\nn
	\times\exp\left\{\frac{1}{8 \lambda} \int\limits_{-\mJ T}^{\mJ T} d u \int\limits_{0}^{\mJ T} dv \, \mathfrak{g}_{\alpha\alpha}(u ,v)\left(-\frac12 \sgn(u)\dd^2_{u}\sgn(u)+\frac{1}{8}\sgn(u)\dd^2_v \sgn(u) - \mJ^2 \e^{g^{cl}_{\alpha\alpha}(u)}\right)\mathfrak{g}_{\alpha\alpha}(u,v)\right\}=\\
	&&\nn =\text{const}\cdot\e^{-I_{\text{on-shell}}}\frac{1}{\det\left(\partial_{u}^2-\frac{1}{4}\partial^2_{v}\right)}\prod_{\alpha=L,R}\left[\det\left(\frac12 \sgn(u)\dd^2_{u}\sgn(u)-\frac{1}{8}\sgn(u)\dd^2_{v}\sgn(v) + \mJ^2 \e^{g^{cl}_{\alpha\alpha}(u)}\right)\right]^{-\frac12},
\eea
where $I_{\text{on-shell}}$ is given by~\eqref{slope-action}, and the constant depends on $\lambda$ exclusively. There are separate determinants for diagonal modes $\fg_{LL}$, $\fg_{RR}$ and for the offdiagonal modes $\fg_{LR}$ and $\fg_{RL}$. The offdiagonal modes are free, so the corresponding determinant is a constant independent of $\mJ T$, as we show in Appendix \ref{sec:1loop-slope-offdiag}. 

To this end, we focus on the determinant of the operator which governs the replica-diagonal modes: 
\begin{equation}
		\mathcal{L} =-\frac{1}{8}\sgn(u)\dd^2_{v}\sgn(u)+2 \sgn(u) a^2\dd^2_{x}\sgn(u) +  \left\{\frac{a}{\cosh \left(\frac{x}{2}\right)}\right\}^{2}\,. \label{mL-slope}
\end{equation}
where we denote $\frac{x}{2}=\tilde{a}\left(\frac{1}{2}-\frac{u}{\mJ T}\right)$, and $\tilde{a}$ is a solution of the constraint (\ref{constr-slope}). The eigenproblem for the operator (\ref{mL-slope}) is similar in a lot of ways to the case of Euclidean large-$q$ SYK at finite temperature \cite{MScomments,Choi19}. In particular, we assume that the eigenfunctions $\Psi_{n,m}(u,v)$ obey the symmetry conditions:
\begin{equation}\label{EF-symmetry}
	\begin{aligned}
		&\Psi_{n,m}(0,v)=0, \\
		&\Psi_{n,m}\left(\mJ T-u, v\pm\frac{\mJ T}{2}\right)=\Psi_{n,m}(u, v)\,.
	\end{aligned}    
\end{equation}
Separation of variables together with these conditions gives the following eigenfunctions\footnote{We follow notations analogous to \cite{Choi19}.}
\begin{equation}
	\begin{aligned}
		&\mathcal{L}\Psi_{n,m}(u,v)=\frac{\left(\frac{2\pi n}{\mJ T}\right)^2+4\frac{m^2}{\mJ^2}}{8}\Psi_{n,m}(u,v), \\
		&\Psi_{n,m}(u,v)=\begin{cases}
			\e^{i2\pi n\frac{v}{\mJ T}}\psi_m^e(t)\sgn(t),\quad n \in 2 \mathbb{Z},\quad m \in \mathcal{M}^{e} \,; \\
			\e^{i2\pi n \frac{v}{\mJ T}} \psi_{m}^{o}(t)\sgn(t), \quad n \in 2 \mathbb{Z}+1,\quad m \in \mathcal{M}^{o}\,,
		\end{cases}
	\end{aligned}    
\end{equation}
where $\psi_m^e(t)$ and $\psi_m^o(t)$ are eigenfunctions of $\left[2a^{2} \partial_{x}^{2}+\frac{a^{2}}{\cosh ^{2}\left(\frac{x}{2}\right)}\right]$ with eigenvalue $\frac{m^2}{2}$ that under $t\rightarrow T-t$ are even and odd respectively\footnote{For the eigenfunctions $\psi$, it is convenient for us to go back to the dimensional time $t = u / \mJ$.}. Note that, generally speaking, this eigenvalue is complex-valued, unlike the Euclidean case \cite{MScomments,Choi19,Lekner}. The explicit form of the eigenfunctions is the following:
\begin{equation}\label{eigenfunctions}
	\begin{aligned}
		\psi_{m}^{e}(t)&=\frac{m}{a} \cosh \left[m\left(\frac{T}{2}-t\right)\right]-\sinh \left[m\left(\frac{T}{2}-t\right)\right] \tanh \left[a\left(\frac{T}{2}-t\right)\right]\,; \\
		\psi_{m}^{o}(t)&=\frac{m}{a} \sinh \left[m\left(\frac{T}{2}-t\right)\right]-\cosh \left[m\left(\frac{T}{2}-t\right)\right] \tanh \left[a\left(\frac{T}{2}-t\right)\right]\,.
	\end{aligned}
\end{equation}
The variable $m$ is fixed by the Dirichlet boundary condition $\psi_{m}^{e, o}(0)=0$. Note that we have to avoid double counting since $\psi_{-m}^{e}(t)=\psi_{m}^{e}(t), \psi_{-m}^{o}(t)=-\psi_{m}^{o}(t)$ \cite{Choi19}. The index sets $\mathcal{M}^{e,o}$ are explicitly defined as follows:
\bea
		\mathcal{M}^{e}&:& \psi_{m}^{e}(0)=0 \Leftrightarrow m \coth \left(\frac{m T}{2}\right)-a \tanh \left(\frac{a T}{2}\right)=0\,; \label{M^e} \\
		\mathcal{M}^{o}&:& \psi_{m}^{o}(0)=0 \Leftrightarrow m \tanh \left(\frac{m T}{2}\right)-a \tanh \left(\frac{a T}{2}\right)=0\,. \label{M^o}
\eea
Hence we can write down the determinant as the specific product over eigenvalues:
\begin{equation}
	\frac{1}{\det(\mathcal{L})}=\prod\limits_{\substack{n\in 2\mathbb{Z}\\ m\in\mathcal{M}^e}}\prod\limits_{\substack{n\in 2\mathbb{Z}+1\\ m\in\mathcal{M}^o}}\frac{2}{\left(\frac{\pi n}{\mJ T}\right)^2+\frac{m^2}{\mJ^2}}. \label{detL-slope}
\end{equation}
Let us find this determinant in the limit of late times $\mJ T\rightarrow\infty$. We can use the formula (\ref{bigJT-a}) to determine $a = \tilde{a} / T$. In this case $\tilde{a}=i(2l+1) \pi+O((\mJ T)^{-1})$, where $l \in \ZZ$. Substituting it into (\ref{M^e})-(\ref{M^o}), we readily find the allowed values of $m$:
\begin{equation}
	\begin{aligned}
		\mathcal{M}^{e}&: m\simeq \frac{i\pi 2k}{T},\quad k\in\mathbb{Z}_+ \\
		\mathcal{M}^{o}&: m\simeq \frac{i\pi (2k-1)}{T},\quad k\in\mathbb{Z}_+,
	\end{aligned}
\end{equation}
where we require $k>0$ to avoid double counting, as mentioned above. Thus the eigenvalues $\frac{m^2}{2}$ are purely negative at late times. Substituting this into the determinant (\ref{detL-slope}) and performing the zeta-function regularization of the products, we finally arrive at the result for the one-loop correction to the disconnected spectral form factor:
\begin{equation}
	\frac{1}{\det(\mathcal{L})}=\prod\limits_{\substack{n\in 2\mathbb{Z}\\ k\in 2\mathbb{Z}_+}}\;\prod\limits_{\substack{n\in 2\mathbb{Z}+1\\ k\in (2\mathbb{Z}+1)_+}}\left(\frac{\mJ T}{\pi}\right)^2\frac{2}{n^2-k^2}=\text{const}\cdot (\mJ T)^{6\sum\limits_{n=1}^{+\infty}1}=\frac{\text{const}}{(\mJ T)^3}.
\end{equation}
This result precisely matches the prediction from the Schwarzian mode, or the triple-scaled limit of SYK \cite{Cotler16,Garcia-Garcia16,Gharibyan18}. We now can improve our semiclassical late-time result (\ref{slope-semiclassical}) with the one-loop correction: 
\begin{equation}
	S(T)_{\text{slope}} = \frac{|\langle Z(iT) \rangle|^2}{2^N} \sim \frac{1}{(\mJ T)^3} \cos^2\frac{\mJ T}{\lambda} \e^{-\frac{\pi^2}{\lambda}}\,. \label{slope-result}
\end{equation}
This formula captures the main properties of the slope region of the spectral form factor in SYK at late times. The decay with the power of $-3$ is caused by the one-loop quantum correction, and the mild oscillations during the decay are caused by the interplay between the complex saddle points and spontaneous breaking of the replica symmetry. As a side remark, our computation shows that for all subleading saddle points on the slope the time decay of the one-loop correction will be the same, so those saddles will remain subleading at all times. 

\section{The ramp region}
\label{sec:Ramp}

\subsection{Large $q$ ansatz}

Now let us proceed to study of the connected spectral form factor. The corresponding field configurations contributing to the path integral (\ref{S(T)-bilocal}) are replica-nondiagonal. So, as we discussed in section \ref{sec:largeQ}, we want to look for replica-nondiagonal solutions of the saddle point equations (\ref{EOM}). Again, to begin let us assume that we are in the regime where the $1/q$-expansion (\ref{1/q-expansion}) is applicable. This time we look for a replica-nondiagonal solution using the ansatz (\ref{G^0-ramp}) and solve the equations of motion (\ref{EOM-ramp-diag})-(\ref{EOM-ramp-offdiag}). The expansion (\ref{1/q-expansion}) can be written as the following ansatz for the fields:
\bea \label{ansatz}
	G_{\alpha\alpha}(t_1, t_2) &=& \frac12 \sgn(t_1 - t_2) \left(1 + \frac{g_{\alpha\alpha} (t_1, t_2)}{q} + o\left(\frac1q\right) \right)\,, \\
	G_{LR}(t_1, t_2) &=& \frac{i}{2} \left(1 + \frac{g_{LR} (t_1, t_2)}{q} + o\left(\frac1q\right) \right)\,,\\
	G_{RL}(t_1, t_2) &=& -\frac{i}{2} \left(1 + \frac{g_{RL} (t_1, t_2)}{q} +  o\left(\frac1q\right) \right)\,; \\
	\Sigma_{\alpha\beta}(t_1, t_2) &=& \frac{\Sigma^{(1)}_{\alpha\beta}(t_1,t_2)}{q} +  o\left(\frac1q\right)\,.
\eea
The antiperiodicity conditions for $G$ imply periodicity conditions for $g$ as follows: 
\begin{equation}\label{conditions}
	\begin{aligned}
		&G_{\alpha\beta}(t_1,t_2)=-G_{\beta\alpha}(t_2,t_1)\Rightarrow g_{\alpha\beta}(t_1,t_2)=g_{\beta\alpha}(t_2,t_1); \\
		&G_{\alpha\beta}(t_1,t_2+T)=-G_{\alpha\beta}(t_1,t_2)\Rightarrow g_{\alpha\beta}(t_1,t_2+T)=g_{\alpha\beta}(t_1,t_2); \\
		&G_{\alpha\beta}(t_1+T,t_2)=-G_{\alpha\beta}(t_1,t_2)\Rightarrow g_{\alpha\beta}(t_1+T,t_2)=g_{\alpha\beta}(t_1,t_2),
	\end{aligned}
\end{equation}
keeping in mind the antiperiodic continuation of the piece-wise constant functions $G^{(0)}$. 

Let us derive the effective action for $g$ on the ramp by substituting these expansion into the action (\ref{action}) and extracting the leading nontrivial $q$-dependence. The Pfaffian term reads 
\be \label{T1L}
\begin{aligned}
	T_1 &:= -\frac12 \Tr \log [\delta_{\alpha\beta} \dd_t - \hat{\Sigma}_{\alpha\beta}] =  -\Tr \log (\dd_t) + \frac12 \Tr \frac{1}{q} (\hat{G}_f \cdot \hat{\Sigma}^{(1)}_{\alpha\beta})+\\&+ \frac12 \Tr \frac{1}{2q^2} (\hat{G}_f \cdot \hat{\Sigma}^{(1)}_{\alpha\gamma} \cdot \hat{G}_f \cdot \hat{\Sigma}^{(1)}_{\gamma\beta}) + O\left(\frac{1}{q^3}\right)\,. 
\end{aligned}
\ee
The polynomial part of the action reads 
\be \label{T2L}
\begin{aligned}
	T_2 &:= \frac{1}{2}\int_0^T \int_0^T dt_1 dt_2  \left(\Sigma_{\alpha\beta}(t_1, t_2) G_{\alpha\beta}(t_1, t_2) - \frac{2^{q-1} \mJ^2}{q^2}s_{\alpha\beta}  G_{\alpha\beta}(t_1, t_2)^q\right) = \\&= 
	\frac{1}{2}\int_0^T \int_0^T dt_1 dt_2 \frac{1}{q} \Sigma^{(1)}_{\alpha\alpha}(t_1, t_2) G_f(t_1, t_2) + \frac{1}{2}\int_0^T \int_0^T dt_1 dt_2 \frac{1}{q^2} \Sigma^{(1)}_{\alpha\alpha}(t_1, t_2) G_f(t_1, t_2) g_{\alpha\alpha} (t_1, t_2)+ \\
	&+\frac12 \int_0^T \int_0^T dt_1 dt_2 \frac{2^{q-1} \mJ^2}{q^2} \frac{1}{2^q}\left(1 + \frac{g_{\alpha\alpha}(t_1, t_2)}{q}\right)^q+\frac{i}{4}\int_0^T \int_0^T dt_1 dt_2 \frac{1}{q} \left[\Sigma^{(1)}_{LR}(t_1, t_2)-\Sigma^{(1)}_{RL}(t_1, t_2)\right]+\\ &+\frac{i}{4}\int_0^T \int_0^T dt_1 dt_2 \frac{1}{q^2} \left[\Sigma^{(1)}_{LR}(t_1, t_2) g_{LR}(t_1, t_2)-\Sigma^{(1)}_{RL}(t_1, t_2) g_{RL}(t_1, t_2)\right]-\\
	&- \frac12 \sum\limits_{\alpha\neq\beta}\int_0^T \int_0^T dt_1 dt_2 \frac{2^{q-1} \mJ^2}{q^2} \frac{1}{2^q}\left(1 + \frac{g_{\alpha\beta}(t_1, t_2)}{q}\right)^q+ o\left(\frac{1}{q^2}\right)\,.
\end{aligned}
\ee
Consider the $1/q$ term. Unlike the slope case, it does not cancel out immediately. We have to integrate out the $\Sigma^{(1)}$ first. After that, the $1/q$-term becomes:
\be \label{BoundaryTerm}
\begin{aligned}
	&\frac{i}{4}\int_0^T \int_0^T dt_1 dt_2 \frac{1}{q} \left[\Sigma^{(1)}_{LR}(t_1, t_2)-\Sigma^{(1)}_{RL}(t_1, t_2)\right]\to \frac{1}{8q}\int_0^T \int_0^T dt_1 dt_2\left[\dd_{t_1}\dd_{t_2}g_{LR}+\dd_{t_1}\dd_{t_2}g_{RL}\right]\\&=\frac{1}{4q}\left[g_{LR}(T,T)-g_{LR}(T,0)-g_{LR}(0,T)+g_{LR}(0,0)\right]=0\,.
\end{aligned}
\ee
We see that this term vanishes due to the periodicity conditions (\ref{conditions}). 

The resulting action in the double-scaling limit reads 
\be \label{Ids[g]L}
\begin{aligned}
	I_{\text{DS}}[g] &= \frac{1}{4 \lambda} \sum_{\alpha=L,R} \int dt_1 dt_2 \left(\frac14 \dd_{t_1}\left(\sgn(t_1-t_2) g_{\alpha\alpha}(t_1, t_2)\right) \dd_{t_2}\left(\sgn(t_1-t_2) g_{\alpha\alpha} (t_1, t_2)\right) + \mJ^2 \e^{g_{\alpha\alpha}(t_1, t_2)}\right)\\
	&-\frac{1}{4 \lambda}\sum\limits_{\alpha\neq\beta} \int dt_1 dt_2 \left(\frac14 \dd_{t_1} g_{\alpha\beta}(t_1, t_2) \dd_{t_2} g_{\alpha\beta}(t_1, t_2) + \mJ^2 \e^{g_{\alpha\beta}(t_1, t_2)}\right)\,. 
\end{aligned}
\ee
The saddle point equations for this action are given by (\ref{EOM-ramp-diag})-(\ref{EOM-ramp-offdiag}). The key distinction from the slope case is that now the offdiagonal components of $g$ have nontrivial potential.

\subsection{General solution at early times}

Like in the slope case, we look for translation-invariant solutions which depend on $t = t_1-t_2$. Then the saddle point equations (\ref{EOM-ramp-diag})-(\ref{EOM-ramp-offdiag}) read

\begin{equation} \label{Liouv}
	\begin{aligned}
		&\partial_{t}^2\left(\sgn(t)g_{\alpha\alpha}(t)\right)=-2\mJ^2\sgn(t) \e^{g_{\alpha\alpha}(t)}, \\
		&\partial_{t}^2g_{\alpha\beta}(t)=-2\mJ^2 \e^{g_{\alpha\beta}(t)},\quad \alpha\neq\beta.
	\end{aligned}
\end{equation}
The general solution for the diagonal components of $g$ is the same as (\ref{genSol-diag}):
\begin{equation} \label{diagSol}
 e^{g_{LL}(t)}=\frac{a_{LL}^2}{\mathcal{J}^{2}\cosh^2(a_{LL}|t|+b_{LL})}\,; \quad
	e^{g_{RR}(t)}=\frac{a_{RR}^2}{\mathcal{J}^{2}\cosh^2(a_{RR}|t|+b_{RR})}\,.
\end{equation}
For the offdiagonal components the general solution reads:
\begin{equation} \label{offdiagSol}
	e^{g_{LR}(t)}=\frac{a_{LR}^2}{\mathcal{J}^{2}\cosh^2(a_{LR} t+b_{LR})}\,; \quad
	e^{g_{RL}(t)}=\frac{a_{RL}^2}{\mathcal{J}^{2}\cosh^2(a_{RL} t +b_{RL})}\,.
\end{equation}
Here $a_{\alpha\beta}$ and $b_{\alpha\beta}$ are complex-valued parameters, generally speaking. Just like in the slope case, we look for a solution on the segment $t \in [0, T/2]$ with specific boundary conditions, then use the property (\ref{G-cont}) to continue it to $t \in [0, T]$. We again will impose the Dirichlet condition on the diagonal components
\be \label{conds}
g_{\alpha\alpha}(0)=0\,,
\ee
which is motivated by the UV behavior. This condition implies the relation
\be \label{g(0)=0}
a_{\alpha\alpha} =\mathcal{J} \cosh b_{\alpha\alpha}. 
\ee
This is the analogue of the constraint (\ref{constr-slope-sq}) for the slope saddle. Note that we are absorbing the sign ambiguity of $a$ into the sign ambiguity of $b$, which we can do because these parameters in (\ref{diagSol}),(\ref{offdiagSol}) are together under $\cosh$.

As far as the offdiagonal components are concerned, let us recall the condition (\ref{G-cont}). Assuming that $\mJ T \ll q$, there is no nontrivial solution of the form (\ref{offdiagSol}) that would satisfy (\ref{G-cont}) (or, equivalently, (\ref{conditions})) and be smooth at $t = T/2$. The technical reason for this is that the constant term $G^{(0)}$ cannot be smoothly continued according to (\ref{G-cont}). Therefore that the smooth replica-nondiagonal solution that would satisfy (\ref{conditions}) does not exist for small times $\mJ T \ll q$ in the framework of perturbative $1/q$-expansion. 

This has a qualitative explanation: from the conditions (\ref{conditions}) it follows that the replica-offdiagonal components $G_{LR}$ must reach zero at some $t_* \in (0, T)$. This means that in terms of the large-$q$ ansatz (\ref{ansatz}), the value of the function $g_{LR}$ must be comparable to $q$ at this finite (compared to $T$) point $t_*$, and hence the $1/q$-expansion (\ref{1/q-expansion}) must break down at this time scale. Thus in order to get a valid replica-nondiagonal saddle point, we have to go beyond the early time regime and $1/q$-expansion into the late times. 

\subsection{Late-time solution}
\label{sec:RampLate}

We again will extrapolate the large $q$ approximation to late times using similar method to the one employed in \cite{Maldacena18,Nosaka19}. Here it has more substance to it than in the slope case discussed in section \ref{sec:SlopeLate}. From the second equation in \eqref{EOM} it follows that $\Sigma_{\alpha\beta}$ varies $q$ times faster than $G_{\alpha\beta}$. Therefore at long times we can approximate $\Sigma_{LR}$ as a delta function:
\begin{equation}\label{RampSigma}
	\Sigma_{L R}(t) =-\Sigma_{RL}(-t)\simeq-i \nu \delta(t), \quad \nu \equiv i \int_{-\infty}^{\infty} d t \Sigma_{L R}=\int_{-\infty}^{\infty} d t\frac{\mJ^2}{q}e^{g_{LR}}=\frac{2 a_{LR}}{q}\sgn(\Re\ a_{LR}).
\end{equation}
The constant $\nu$ is determined from the short-time solution for $\Sigma_{LR}$ (\ref{offdiagSol}).   

$\Sigma_{LL}$ and $\Sigma_{RR}$ are odd functions of $t$, leading to a $\delta'(t)$ with some coefficient that is proportional to $1/q$. In the first equation in~\eqref{EOM} there is already $\partial_t G$ which is of order zero in $q$, so we can neglect the terms containing $\Sigma_{LL}$ and $\Sigma_{RR}$. Thus at very long times the equations~\eqref{EOM} reduce to 
\begin{equation}
		\begin{aligned}
	&\dd_t G_{LL}+i\nu G_{RL}=0\,; \\
	&\dd_t G_{LR}+i\nu G_{RR}=0\,; \\
	&\dd_t G_{RR}-i\nu G_{LR}=0\,; \\
	&\dd_t G_{RL}-i\nu G_{LL}=0\,.
\end{aligned}
\end{equation}
Taking into account the fact that $G_{\alpha\alpha}$ should be symmetric around $T/2$, we can write the solution as 
\begin{equation} 
	\begin{aligned}
		G_{L L}&=A \cosh [\nu(T / 2-t)], \quad G_{RL}=-i A \sinh [\nu(T / 2-t)],\\
		G_{RR}&=B \cosh [\nu(T / 2-t)], \quad G_{LR}=i B \sinh [\nu(T / 2-t)].
	\end{aligned}
\end{equation}
But $G_{LR}(T/2-t)=G_{RL}(T/2+t)$, so $A=B$ and finally we get:
\begin{equation} \label{largeTimeSol}
	G_{L L}=G_{RR}=A \cosh [\nu(T / 2-t)], \quad G_{RL}=-G_{LR}=-i A \sinh [\nu(T / 2-t)].
\end{equation}
We are left with a free parameter $A$ to be determined by gluing to the short-time region. The exact mapping from the short-time parameter $a$ or $b$ to $A$ will depend on the specific way $T$ scales with $q$. We proceed to analyze these specific regimes. 

Expanding ~\eqref{diagSol},~\eqref{offdiagSol} at late times and~\eqref{largeTimeSol} at early times gives
\begin{equation}\label{asympt}
	\begin{aligned}
		G_{L L} & \sim \frac{1}{2}\left(1+\frac{g_{L L}}{q}\right) \sim \frac{1}{2}-\frac{1}{q}\left(\frac{1}{2}\log \left(\frac{\mJ}{2a_{LL}}\right)^2+b_{LL}+a_{LL} t\right)=A \cosh \frac{\nu T}{2}-\nu t A \sinh \frac{\nu T}{2}, \\
		i G_{RL} & \sim \frac{1}{2}\left(1+\frac{g_{RL}}{q}\right) \sim \frac{1}{2}-\frac{1}{q}\left(\frac{1}{2}\log \left(\frac{\mJ}{2a_{RL}}\right)^2+b_{RL}+a_{RL}t\right)=A \sinh \frac{\nu T}{2}-\nu t A \cosh \frac{\nu T}{2},
	\end{aligned}
\end{equation}
and similar conditions for $G_{RR}$ and $G_{LR}$. These equations are consistent only in the limit $\Re\ \nu T\rightarrow\infty$.  Taking this limit, the above equations imply
\begin{equation} \label{asymp1}
	\begin{aligned}
		&\frac{1}{2}A\e^{\frac{\nu T}{2}}=\frac{1}{2}\Rightarrow A=\e^{-\frac{\nu T}{2}}, \\
		&\frac{1}{2}\nu A\e^{\frac{\nu T}{2}}=\frac{a_{LL}}{q}=\frac{a_{RL}}{q}=\frac{a_{LR}}{q}=\frac{a_{RR}}{q}\Rightarrow a_{LL}=a_{RR}=a_{RL}=a_{LR},\, \Re\ a_{LR}>0.
	\end{aligned}
\end{equation}
In particular, this means that we can now take off replica indices from the parameter $a$. Next, we have the relation
\begin{equation}
	A\cosh \frac{\nu T}{2}+\frac{b_{LL}}{q}=A\sinh \frac{\nu T}{2}+\frac{b_{RL}}{q}\,. 
\end{equation}
Taking into account the above expression for $A$, we notice that it only makes sense in the late time limit as 
\bea
b_{RL}&=&b_{LL}+\sigma\,;\label{asymp2}\\
	b_{LL}&=&b_{RR},\quad b_{LR}=b_{RL}\,, \label{asymp3}
\eea
where $\sigma$ is defined as
\begin{equation} \label{sigma}
	\frac{\sigma}{q}=e^{-\nu T}\,.
\end{equation} 

\paragraph{Times of order $\mJ T\sim q\log q$.} In the time regime we scale the time so that $\sigma$ is a finite constant independent of $q$ or $T$. This defines the scaling regime for the time $\mJ T$ with $q$ for which the constructed solutions exists globally. However, since $\nu$ is generally complex-valued, we would like to isolate the real scaling variable. The relation (\ref{sigma}) can be rewritten as 
\bea
| \sigma|=q e^{-\Re\ \nu T}\,;\\
\arg\ \sigma = - \Im\ \nu T\,. \label{arg-sigma}
\eea
The $|\sigma|$ fixes the scaling of the time with large $a$: 
\be
T = \frac{1}{\Re\ \nu} \log \frac{q}{|\sigma|}\,.
\ee
Substituting it into (\ref{arg-sigma}), we get that 
\be
\arg\ \sigma = \tan (\arg \nu) \log \frac{q}{|\sigma|}\,.
\ee
Since $\frac{\Im\ \nu}{\Re\ \nu}=\frac{\Im\ a}{\Re\ a}$ is independent of $q$, we see that $\sigma$ as defined by (\ref{sigma}) has a phase that grows as $\log q$ if $\Im\ a \neq 0$. Because of the relation (\ref{asymp2}), this would mean that in this case the solution also does not have a well-defined large $q$ limit. Thus we conclude that $\nu$ and, consecutively, $a$ must be real-valued. Taking into account the relation $a = \mJ \cosh b$ and the gluing conditions (\ref{asympt}), we see that the free parameter $b_{LL}$ has to be real as well.  The $\sigma$ remains another free real-valued parameter.

\begin{figure}[t]
	\center{\includegraphics[scale=0.35]{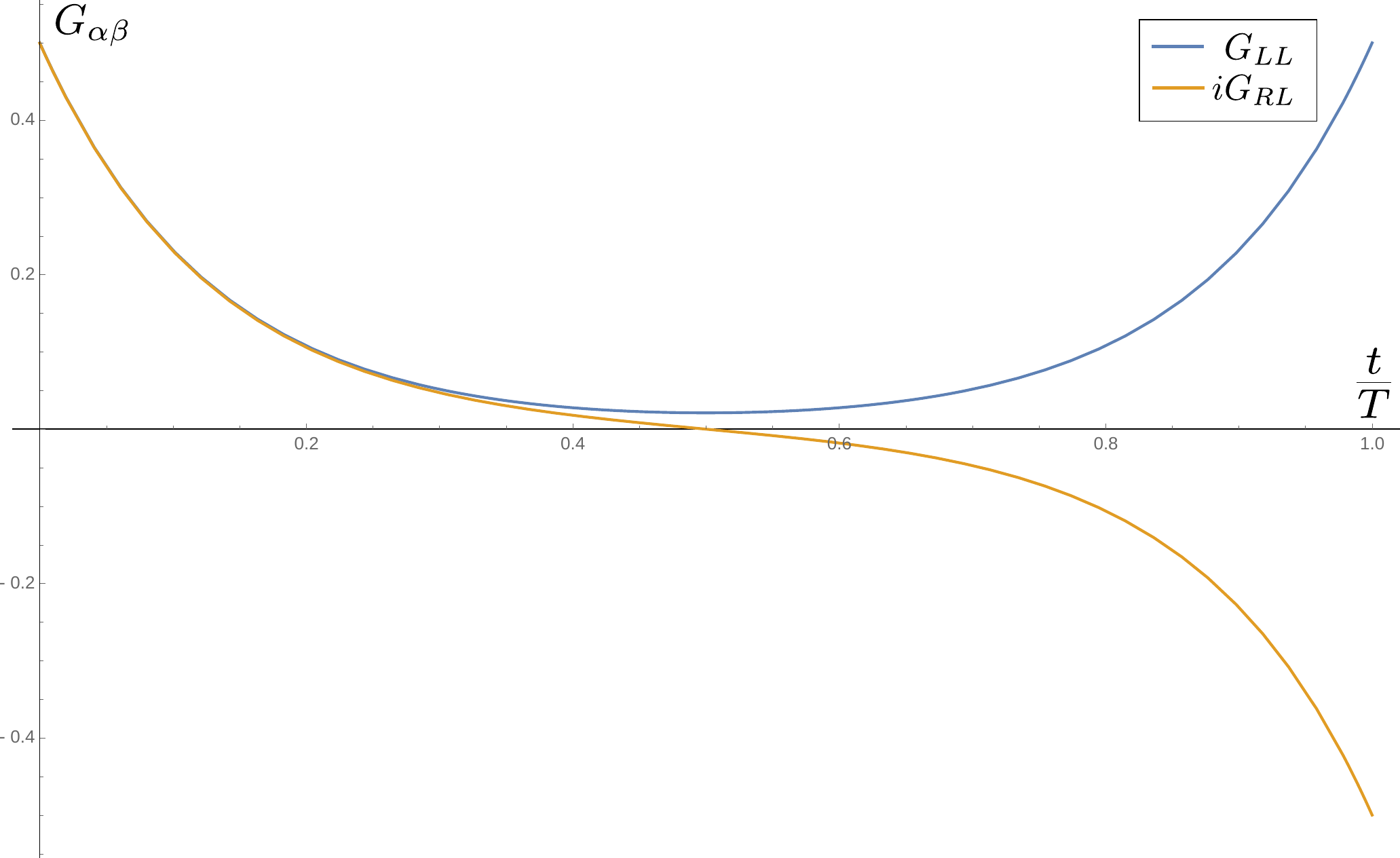}}\caption{Solution for the ramp saddle point glued together from the early-time part (\ref{smallTSol-diag})-(\ref{smallTSol-offdiag}) and the late-time part (\ref{qlogqSol}).}\label{fig:DiagANDNonDiag}
\end{figure}

Thus in the present regime the above asymptotic analysis reduces the parameter space of the solutions to two independent parameters $b_{LL}$ and $\sigma$. In what follows we omit index $LL$ in $b_{LL}$. Substituting the relations (\ref{asymp1}),(\ref{asymp2}),(\ref{asymp3}) back into the general solutions (\ref{diagSol}),(\ref{offdiagSol}), the resulting small time solution reads
\bea \label{smallTSol-diag}
	e^{g_{RR}(t)} &=& e^{g_{LL}(t)}=
		\frac{\cosh^2 b}{\cosh^2(\mJ(\cosh b) |t|+b)}\,; \\
 e^{g_{LR}(t)}&=& e^{g_{RL}(t)}= \label{smallTSol-offdiag}
		\frac{\cosh^2 b}{\cosh^2(\mJ(\cosh b) t+b+\sigma)}\,.
\eea
The resulting late-time solution is
\begin{equation} \label{qlogqSol}
	G_{RR}=G_{L L}=\e^{-\frac{\nu T}{2}} \cosh [\nu(T / 2-t)], \quad G_{RL}=-G_{LR}=-i \e^{-\frac{\nu T}{2}} \sinh [\nu(T / 2-t)],\quad \nu=\frac{2a}{q}.
\end{equation}
The composite solution~\eqref{smallTSol-diag},(\ref{smallTSol-offdiag})+\eqref{qlogqSol} is smooth on the interval $[0,T]$, as we glued their asymptotics up to the term $\sim t$. The composite solution is schematically plotted on the Fig. \ref{fig:DiagANDNonDiag}. 

\paragraph{Times of order $\mJ T\sim q^\alpha$, $0<\alpha\leq1$.} This is the limiting case of the above regime, where $\sigma\sim q\rightarrow\infty$, so $b_{RL}\sim q\rightarrow\infty$. This immediately implies $|2iG_{\alpha\beta}(t_1,t_2)|<1$, and from the second saddle point equations (\ref{EOM}) $\Sigma_{RL}=i\frac{\mJ^2}{q} (2iG_{RL}(t_1,t_2))^{q-1}\rightarrow 0$. Thus, the $1/q$ expansion for non-diagonal Green functions breaks and we come to the large $t$ equations:
\begin{equation}
	\partial_t G_{\alpha\beta}(t)=0\,.
\end{equation}
These are the exact same equations as in the slope case, see (\ref{EOM-slope-late}), up to a constant. Thus, repeating the previous argument, we conclude that in this region there is no replica-nondiagonal solution. 

\subsection{On-shell action and symmetries}
\label{sec:RampAction}

Now let us compute the on-shell action. Since the replica-nondiagonal solution exists beyond the $1/q$-expansion, we need to use the full action (\ref{action}). For translationally invariant solutions it reads
\begin{equation}
	\frac{I[G, \Sigma]}{N} = -\log \Pf[\delta_{\alpha\beta} \dd_t - \hat{\Sigma}_{\alpha\beta}] +\frac{T}{2}\int_{-T}^T dt \left(\Sigma_{\alpha\beta}(t) G_{\alpha\beta}(t) - \frac{2^{q-1} \mJ^2}{q^2}s_{\alpha\beta} G_{\alpha\beta}(t)^q\right)\,.
\end{equation}
We assume that $\mJ T$ is very large. Strictly speaking, we should separate the time integration into early time segment $[0, \tau]$ and late time segment $[\tau, T]$. The solution on the former is given by (\ref{smallTSol-diag})-(\ref{smallTSol-offdiag}), and the solution on the latter is given by (\ref{qlogqSol}), with $\Sigma$ given by the delta-function as in (\ref{RampSigma}). The gluing time scale $\tau$ should be of order $q \log q$. However, the solution (\ref{smallTSol-diag}) exponentially decays with time. So the integrals of the type $\int_0^\tau dt \Sigma G$ behave as $\frac{1}{q}\e^{-\kappa \tau}$. Therefore, the early-time solution makes an exponentially small contribution to the action at late times. Hence we can neglect it and use the long-time solution (\ref{qlogqSol}) for all $t \in [0, T]$, with $\Sigma_{LR}$ given by delta-function as (\ref{RampSigma}) and $\Sigma_{LL}=\Sigma_{RR}=0$.

Let us start with the Pfaffian term. Expanding the Pfaffian into series, we separate the free term, replacing $\Tr \log \dd_t \to \log 2$ so that $Z(0) = 2^{\frac{N}{2}}$ holds. We evaluate the rest of the series directly (summation over replica indices is implicit): 
\begin{equation}
	\begin{aligned}
		&-\frac{1}{2}\Tr\log [\delta_{\alpha\beta} \dd_t - \hat{\Sigma}_{\alpha\beta}]=-\Tr \log (\dd_t) + \frac12 \Tr (\hat{G}_f \cdot \hat{\Sigma}_{\alpha\beta})+\frac14 \Tr (\hat{G}_f \cdot \hat{\Sigma}_{\alpha\gamma} \cdot \hat{G}_f \cdot \hat{\Sigma}_{\gamma\beta}) +\dots=\\
		&=-\log2-\frac{(\nu T)^2}{8}+\frac{(\nu T)^4}{192}-\frac{(\nu T)^6}{2880}+\dots=-\log2-\log\cosh\frac{\nu T}{2}=-\frac{\nu T}{2}+O(\e^{-\nu T}).
	\end{aligned}
\end{equation}
Evaluating the polynomial term in the action, we get:
\begin{equation}
	\frac{T}{2}\left(1-\frac{1}{q}\right)\int_{-T}^T dt\,\Sigma_{\alpha\beta}(t) G_{\alpha\beta}(t)=\frac{\nu T}{2}\left(1-\frac{1}{q}\right)=\frac{\nu T}{2}\left(1-\frac{1}{\sigma}\e^{-\nu T}\right).
\end{equation}
We see that the linear in $T$ terms cancel out from the total action, so the resulting classical action on the ramp is
\begin{equation}
	I_{\text{ramp}}= 0 + O(\nu T\,\e^{-\nu T}).
\end{equation}
This matches the result obtained in \cite{Saad18} for numerical finite-$q$ solutions on the ramp, and is the main reason for the linear growth on the ramp.  

One particular corollary is that all ramp solutions have the same action, so to compute the spectral form factor on the ramp, we need to integrate over the parameter space. To establish it, let us recollect the symmetries that are spontaneously broken by the solutions (\ref{smallTSol-diag})-(\ref{smallTSol-offdiag}). 
\begin{itemize}
	\item $b \rightarrow - b$. We take care of this symmetry by integrating over $b \in (-\infty, +\infty)$. 
	\item Parity symmetry $G_{LR}(t_1, t_2) \to -G_{LR}(t_1, t_2)$. In terms of the early-time solution, it reduces to the replacement $G_{LR}^{(0)} = \frac{i}{2} \to -\frac{i}{2}$. To account for this symmetry, we add a factor of $2$ into the spectral form factor. 
	\item Spontaneously broken time translations. The replica-diagonal solutions (\ref{smallTSol-diag})-(\ref{smallTSol-offdiag}) break time translations in both replicas as $U(1) \times U(1) \to U(1)$, as is commonplace for such solutions \cite{AKTV,AKV,Saad18}. In the obtained solutions, the parameter $\sigma$ plays the role of the time translation zero mode. To see this, let us introduce new parameter 
	\be
	\Delta = \frac{\sigma}{\mJ \cosh b}\,.
	\ee
	Then the solution for the offdiagonal component (\ref{smallTSol-offdiag}) reads: 
	\bea 
	e^{g_{LR}(t)}&=& e^{g_{RL}(t)}= \label{smallTSol-offdiag-delta}
	\frac{\cosh^2 b}{\cosh^2(\mJ(\cosh b) (t+\Delta)+b)}, 
	\eea
	So keeping in mind the antiperiodicity, we have to integrate over $\Delta\in [0, T]$ to sum over the orbit of the broken symmetry. 
\end{itemize}
Now we almost have all of the ingredients to compute the ramp spectral form factor at late times. The only missing piece is the one-loop correction, which we deal with in the next section.

\subsection{One-loop correction}
\label{sec:RampQuantum}

To study the one-loop correction, we use the same approach as in section \ref{sec:1loop-slope} for the slope. In the appendix \ref{sec:RampQuantum-deriv} we derive the quantum correction from the full action (\ref{action}) and show that all of the nontrivial quantum dynamics come from the early times. So we can work directly in terms of the $g$ quantum variable. We split $g_{\alpha\beta}$ into classical and quantum parts:
\begin{equation}
	g_{\alpha\beta}=g_{\alpha\beta}^{cl}+\mathfrak{g}_{\alpha\beta},
\end{equation}
where $g_{\alpha\beta}^{cl}$ for $t \in \left[0, \frac{T}{2}\right]$ are given by in~\eqref{smallTSol-diag},(\ref{smallTSol-offdiag}) and periodically continued to $\left[\frac{T}{2}, T\right]$. Keeping in mind the exponential decay of the classical solution $\e^{g^{\text{cl}}}$, we can approximate the potential term from the full quantum action (\ref{1loop-ramp-full}) as follows: 
\be
\mJ^2 \left[2G^{\text{cl}}_{\alpha\beta}(u)\right]^{q}\left(G^{\text{cl}}_{\alpha\beta}(u)\right)^{-2}s_{\alpha\beta}\mathfrak{G}^2_{\alpha\beta}(u,v) \simeq \frac{\mJ^2}{q^2} \left(\e^{g^{cl}_{\alpha\beta}(u)}+\e^{g^{cl}_{\alpha\beta}(\mJ T-u)}\right)\mathfrak{g}_{\alpha\beta}^2(u,v) \,.
\ee
Alternatively, one can get this potential by extracting the quadratic part of the early-time effective action (\ref{Ids[g]L}).

Using the early time solutions (\ref{smallTSol-diag})-(\ref{smallTSol-offdiag}) for fixed $b$ and $\Delta = \frac{\sigma}{\mJ \cosh b}$, the 1-loop correction to the spectral form factor is given by:
\begin{equation}\label{SFFramp}
	\begin{aligned}
		&S(T) \to \left[\det \left(\frac12 \dd^2_{u}-\frac{1}{8}\dd^2_{v} + \mJ^2 \e^{g^{cl}_{LR}(u)}+ \mJ^2 \e^{g^{cl}_{LR}(\mJ T-u)}\right)\times\right.\\
		&\left.\times\det\left(\frac12 \sgn(u)\dd^2_{u}\sgn(u)-\frac{1}{8}\sgn(u)\dd^2_{v}\sgn(u) + \mJ^2 \e^{g^{cl}_{LL}(u)}+ \mJ^2 \e^{g^{cl}_{LL}(\mJ T-u)}\right)\right]^{-1}.
	\end{aligned}
\end{equation}
Because $g^{\text{cl}}$ decays exponentially at late times, we can approximate the differential operator as a sum of two operators acting for $t < T/2$ and $t > T/2$. Let us consider first the determinant for the diagonal modes, where the potential is determined by $g_{LL}^{\text{cl}}$. 
\begin{equation}
	\begin{aligned}
		\mathcal{L_{\text{diag}}}&=\frac12 \sgn(u)\dd^2_{u}\sgn(u)-\frac{1}{8}\sgn(u)\dd^2_{v}\sgn(u) +  \left\{\frac{a}{\cosh \left(a t+b\right)}\right\}^{2}+  \left\{\frac{a}{\cosh \left(a(T-t)+b\right)}\right\}^{2}\\
		&\simeq\left(-\frac{1}{8}\sgn(u)\dd^2_{v}\sgn(u)+2 \sgn(u)a^2\dd^2_{x}\sgn(u) +  \left\{\frac{a}{\cosh \left(\frac{x}{2}\right)}\right\}^{2}\right)\theta\left(\frac{T}{2}-t\right)+\\
		&+\left(-\frac{1}{8}\sgn(u)\dd^2_{v}\sgn(u)+2 \sgn(u)a^2\dd^2_{\tilde{x}}\sgn(u) +  \left\{\frac{a}{\cosh \left(\frac{\tilde{x}}{2}\right)}\right\}^{2}\right)\theta\left(t-\frac{T}{2}\right),
	\end{aligned}
\end{equation}
where $a=\mJ \cosh b$ and we denoted $\frac{x}{2}=at+b$ and $\frac{\tilde{x}}{2}=aT-at+b$. Let us find its spectrum.

The eigenfunctions $\Psi_{n,m}(u,v)$ observe the symmetry conditions (\ref{EF-symmetry}):
\begin{equation}
	\begin{aligned}
		&\Psi_{n,m}(0,v)=0, \\
		&\Psi_{n,m}\left(\mJ T-u,v\pm\frac{\mJ T}{2}\right)=\Psi_{n,m}(u, v).
	\end{aligned}    
\end{equation}
Proceeding analogously to the slope case with separation of variables, we write 
\begin{equation}
	\begin{aligned}
		&\mathcal{L}_{\text{diag}}\Psi_{n,m}(t,T')=\frac{\left(\frac{2\pi n}{\mJ T}\right)^2+4\frac{m^2}{\mJ^2}}{8}\Psi_{n,m}(t,T'), \\
		&\Psi_{n,m}(t,T')=\begin{cases}
			\e^{i2\pi n\frac{T'}{T}}\psi_m^{e\pm}(t)\sgn(t),\quad n \in 2 \mathbb{Z},\quad m \in \mathcal{M}^{e}\,; \\
			\e^{i2\pi n \frac{T'}{T}} \psi_{m}^{o\pm}(t)\sgn(t), \quad n \in 2 \mathbb{Z}+1,\quad m \in \mathcal{M}^{o} \,,
		\end{cases}
	\end{aligned}    
\label{Ldiag}
\end{equation}
where $\psi_m^e(t)$ and $\psi_m^o(t)$ are eigenfunctions of $\left[2a^{2} \partial_{x}^{2}+\frac{a^{2}}{\cosh ^{2}\left(\frac{x}{2}\right)}\right]\theta\left(\frac{T}{2}-t\right)+\left[2a^{2} \partial_{\tilde{x}}^{2}+\frac{a^{2}}{\cosh ^{2}\left(\frac{\tilde{x}}{2}\right)}\right]\theta\left(t-\frac{T}{2}\right)$ with eigenvalue $\frac{m^2}{2}$ that under $t\rightarrow T-t$ are even and odd respectively. To write down the explicit form of eigenfunctions, let us introduce auxiliary notation:
\bea
P_m(t) &=& \frac{m}{a} \cosh \left(m t+\frac{b}{a}\right)-\sinh \left(m t+\frac{b}{a}\right) \tanh \left(at+b\right)\,;\\
Q_m(t) &=& \frac{m}{a} \sinh \left(m t+\frac{b}{a}\right)-\cosh \left(m t+\frac{b}{a}\right) \tanh \left(at+b\right)\,.
\eea
Then we can write down $\psi_m$ as follows:
\begin{equation}\label{eigenfunctions-ramp}
	\begin{aligned}
		\psi_{m}^{e+}(t)&=P_m(t)\theta\left(\frac{T}{2}-t\right)+P_m(T-t) \theta\left(t-\frac{T}{2}\right)\,; \\
		\psi_{m}^{e-}(t)&=Q_m(t) \theta\left(\frac{T}{2}-t\right)+Q_m(T-t)\theta\left(t-\frac{T}{2}\right), \\\psi_{m}^{o+}(t)&=P_m(t)\theta\left(\frac{T}{2}-t\right)-P_m(T-t)\theta\left(t-\frac{T}{2}\right), \\
		\psi_{m}^{o-}(t)&=Q_m(t)\theta\left(\frac{T}{2}-t\right)-Q_m(T-t)\theta\left(t-\frac{T}{2}\right)\,.
	\end{aligned}
\end{equation}
The sets $\mathcal{M}^{e}=\mathcal{M}^{o}\equiv\mathcal{M}$ ended up being identical. They are defined by the conditions :\begin{equation}
	\begin{aligned}
		&\psi_{m}^{e,o+}(0)=0 \Leftrightarrow m \coth \left(\frac{m b}{a}\right)-a \tanh \left(b\right)=0, \\
		&\psi_{m}^{e,o-}(0)=0 \Leftrightarrow m \tanh \left(\frac{m b}{a}\right)-a \tanh \left(b\right)=0. \label{M-ramp}
	\end{aligned}
\end{equation}
The key point is that these equations have no time dependence, unlike the slope case (\ref{M^e})-(\ref{M^o}). The solutions have a general form\footnote{The $\pm$ labels one equation in (\ref{M-ramp}), and $k$ labels an individual solution.}:
\be
m = \mJ f_{\pm,k}(b)\,,
\ee
with $f_{\pm,k}$ being a complex-valued function. Then the inverse determinant of $\mL$ reads
\be
		\frac{1}{\det(\mathcal{L})}=\prod\limits_{\substack{n\in 2\mathbb{Z}\\ m\in\mathcal{M}}}\prod\limits_{\substack{n\in 2\mathbb{Z}+1 \\s m\in\mathcal{M}}}\frac{2}{\left(\frac{\pi n}{\mJ T}\right)^2+\frac{m^2}{\mJ^2}}=\prod_{l=+,-}\prod\limits_{k}\prod\limits^{n_0}_{n\in \mathbb{Z}}\frac{2}{\left(\frac{\pi n}{\mJ T}\right)^2+f_{l,k}^2(b)}\,. \label{detL-ramp-1}
\ee	
Note that rather than using the zeta-function regularization throughout, we instead introduce a cutoff $n_0$ in the product over $n$. Let us recall that the ramp saddle only exists for $\mJ T \sim q \log q$. For consistency of the computation with the large $q$ limit, we assume that $n_0 \ll q$. This means that we can neglect the first term in the denominator of (\ref{detL-ramp-1}):
\be	
		\frac{1}{\det(\mathcal{L})}\simeq\prod\limits_{k}\prod^{n_0}\limits_{n\in \mathbb{Z}}\frac{2}{f_{+,k}^2(b) f_{-,k}^2(b)}\sim\prod\limits_{k}(f_{+,k}(b)f_{-,k}(b))^{-4\sum_{n=1}^{n_0}1}=\prod_k f_{+,k}^{2}(b)f_{-,k}^{2}(b) =: \mu^{1/2}(b). \label{detL-ramp-2}
\ee
Here we returned to the zeta-function regularization using the relations between the hard cut-off $n_0$ and the zeta-function cut-offs $\Lambda_1$ and $\Lambda_2$:
\begin{equation}\label{regs}
	\zeta(0) = \sum\limits_{n=1}^{n_0}1=n_0=-\frac{1}{2}+\Lambda_1,\text{ and } \zeta'(0) = -\sum\limits_{n=1}^{n_0}\log n=-\log (n_0!)=-\frac{1}{2}\log(2\pi)+\Lambda_2\,,
\end{equation}
and dropped the divergences. 

We see that the one-loop determinant only depends on $b$, in agreement with the results from \cite{Saad18}. Let us know study the asymptotic behavior of $\mu^{1/2}(b)$ as defined by (\ref{detL-ramp-2}). 
\begin{itemize}
	\item $b\to 0$. The equations (\ref{M-ramp}) have the solution
	\be
	f_{+,k} = \frac{i\pi}{2b}k\,, k \in \ZZ_+; \qquad f_- = 1\,.
	\ee
	We assume that $k > 0$ to avoid double counting in eigenfunctions $\psi_m$. Then 
	\be
	\mu^{\frac{1}{2}}(b) = \prod_k \left(-\frac{\pi^2}{4b^2}k^2\right) \simeq  4b\,.
	\ee
	We have eliminated the factor of $-1$ in the product, because it will cancel out in the spectral form factor with the second copy of the determinant. 
	\item $b \to \infty$. The equations (\ref{M-ramp}) in this limit have unique solutions
	\be
	f_{\pm} = \cosh b\,.
	\ee
	Then there is no product in $\mu^{1/2}$, and it yields 
	\be
	\mu^{1/2}(b) = \cosh^4 b\,.
	\ee
\end{itemize}
The function $\mu^{1/2}$ is the result for the contribution of diagonal modes to the total one-loop correction. What about the off-diagonal modes? Because of the spontaneously broken time translation symmetry, from the solutions (\ref{smallTSol-diag})-(\ref{smallTSol-offdiag-delta}) we see that it is true that 
\be
g_{LL}^{\text{cl}} (t) = g_{LR}^{\text{cl}} (t-\Delta)\,.
\ee
This means that the corresponding differential operator $\mL_{\text{offdiag}}$ is the same as the operator $\mL_{\text{diag}}$ up to a simple time shift in the potential. Therefore the eigenvalues and the determinant will be the same. Thus the total one-loop correction to the spectral form factor is given by $\mu^{1/2} (b) \mu^{1/2} (b) = \mu (b)$. 

Now we are finally ready to write down the result for the spectral form factor on the ramp solution with the one-loop correction. Recall that the on-shell action is zero, and thus we are only left with the one-loop correction integrated over the parameters and orbits of broken symmetries, as discussed in section \ref{sec:RampAction}. 
\be
S(T)_{\text{ramp}} = 2 \times 2^{-N} \mJ \int_0^T d\Delta \int_{-\infty}^{+\infty} \mu(b) db = 2 \times 2^{-N} \mJ T \int_{-\infty}^{+\infty} \mu(b) db\,.  \label{SFF-ramp-result}
\ee
Recall that the factor of $2$ comes from the parity symmetry. The factor of $2^{-N}$ comes from the normalization by $Z(0)^2$ in the definition of $S(T)$ (\ref{S(T)}). We have obtained the linear growth on the ramp from analytic solutions. The measure $\mu(b)$ is determined on the one-loop level via the equation (\ref{detL-ramp-2}) and equations (\ref{M-ramp}), and has the following asymptotic properties: 
\bea
b \to 0: \qquad \mu(b) &\sim& b^2 \,;\\
b \to \pm \infty: \qquad \mu(b) &\sim& \cosh^8 b\,.
\eea
Comparing the result with foundings of \cite{Saad18}, we see that the parameter $b$ has the meaning of the effective temperature in terms of the SYK saddles, or size of the double trumpet geometry in the holographic JT gravity description.

\section{Time scales of chaos}
\label{sec:Transition}

In this section we recollect the notable time scales that occurred, assuming the double scaling limit, and discuss the transition from the slope to the ramp. We proceed from the early time scales to late times. 

\paragraph{Replica symmetry breaking time $T_{\text{cr}}$}. This time is determined by the moment when all solutions of the constraint (\ref{constr-slope}) become complex-valued, and pairs of complex-conjugate saddles starts dominating instead of a single saddle for each replica. This leads to the oscillations of the spectral form factor for $T > T_{\text{cr}}$, expressed by the $\cos \frac{\mJ T}{\lambda}$ prefactor in the expression (\ref{slope-result}). In the large $q$ limit, $T_{\text{cr}} \sim O(1)$ is finite. As explained in \cite{Cotler16,Gharibyan18}, these oscillations are generic for SYK but special to the $\beta=0$ case. They originate from contributions of the two edges of the spectrum, which are not smoothed out at $\beta = 0$. 

\paragraph{Thouless time $T_{\text{Th}}$.} This time scale is usually defined as the time of the onset of universal RMT behavior, and in case of the spectral form factor the mentioned universal behavior is the ramp \cite{Altland17,Altland20,Nosaka18,Saad18,Gharibyan18,Garcia-Garcia16,Sonner17}, so sometimes $T_{\text{Th}}$ is also referred to as ramp time. In our case, the Thouless time is defined by the moment when the replica-diagonal saddle point appears. As we discussed in section \ref{sec:Ramp}, this time scale is of order $q \log q$, or more specifically from (\ref{sigma})
	\be
T_{\text{Th}} \sim \frac{q}{2 \mJ \cosh b} \log \frac{q}{\mJ \Delta \cosh b}\,. \label{Thouless}
\ee
We can compare this result with finite-$q$ investigations \cite{Gharibyan18,Altland17,Nosaka18} if we express $q$ through $N$ via $\lambda = \frac{q^2}{N}$:
\be
T_{\text{Th}} \sim \frac{\sqrt{\lambda}}{4 \mJ \cosh b} \sqrt{N} \log N\,. \label{Thouless-N}
\ee
Interestingly, this matches the result of \cite{Altland17} obtained from an effective replica field theory for finite $q$ SYK, but is at odds with the numerical result of \cite{Gharibyan18}. In our case we have obtained this scale by analyzing the consistency conditions of extrapolation of the leading order of the $1/q$-expansion to the late times. It is plausible that higher orders in $1/q$ can decrease this time scale. 

\paragraph{Dip time $T_{\text{dip}}$.} The dip time is defined as the time when the spectral form factor transitions from the slope to the ramp regime, or, in other words, when the replica-nondiagonal ramp saddle point becomes dominant over the replica-diagonal slope saddle point. Equating the slope (\ref{slope-result}) to the ramp (\ref{SFF-ramp-result}), we get (up to constants independent of $N$ or $q$)
\begin{equation}
	2^{-N}\mJ T_{\text{dip}} \sim \e^{-\frac{\pi^2}{\lambda}}\cos^2\frac{\mJ T_{\text{dip}}}{\lambda}\frac{1}{(\mJ T_{\text{dip}})^3}\Rightarrow T_{\text{dip}}\sim \e^{\alpha N},\quad \alpha>0.
\end{equation}
This is in full agreement the numerics for finite-$q$ SYK and generalizations \cite{Cotler16,Gharibyan18,Hunter-Jones17,delCampo17,Sonner17,Garcia-Garcia18,Nosaka19,Gur-Ari18}. Since this is a time scale when the dominant saddle point changes, another effect that is associated with it in the double-scaled SYK is the breakdown of the $1/q$-expansion (\ref{1/q-expansion}) in the spectral form factor, since the replica-diagonal solution $G_{\alpha\beta}(t)$ only exists as its extrapolation in the late-time regime. Thus the slope-ramp transition shows that the $1/q$-expansion breaks down at the time scales of order $ \e^{\gamma q^2}$. 

\section{Discussion and outlook}
\label{sec:Discussion}

The main result of the present work is the analytic description of the slope and ramp regimes of the spectral form factor in the double scaled SYK model. Specifically, 
	\begin{itemize}
	\item We have constructed analytic solutions in the large $q$ SYK which correspond to saddle points contributing to the slope and the ramp regions of the spectral form factor. In terms of the collective field path integral formulation, the slope is described by a discrete set of replica-diagonal solutions parametrized by solutions of the equation (\ref{constr-slope}), and the ramp is described by a continuous set of the replica-nondiagonal solutions parametrized by $b \in \RR$ and $\Delta \in [0, T]$.
	\item We have found that slope solutions exist at all times and are always valid within the perturbative $1/q$-expansion. 
	\item We have shown that ramp solutions only exist for times of order $q \log q$ or later, and we need to go beyond the perturbative $1/q$-expansion to construct them. 
	\item We see that the slope region has a discrete set of subleading saddle points parametrized by integer numbers, analogous to the subleading saddles in finite $q$ SYK \cite{Cotler16,AKTV,AKV}. 
	\item We see that in the ramp regime, all existing replica-nondiagonal saddles, that are invariant with respect to synchronous time translations in both replicas, are dominant and have zero action. 
	\item There is a phase transition on the slope accompanied by the replica symmetry breaking, which generates the oscillations at times of order 1.
	\item We identify the time scale at which the self-consistent replica-nondiagonal solution exists as the Thouless time, and we have also evaluated the dip time. 
\end{itemize}
As our study is limited to the double-scaled limit of SYK, there are open questions.

\textbf{1.} The plateau regime is inaccessible to the path integral (\ref{S(T)-bilocal}). As pointed out in \cite{Saad18,Cotler16} and confirmed in \cite{Altland20,SSS,Saad19,Nayak19}, the plateau in regular SYK is described by the doubly-nonperturbative effects, which are of order $\e^{-\e^N}$. However it can be treated analytically by certain effective field theories, e.g. \cite{Altland20,Altland17}. It would be interesting to see if there is a replacement for the double-scaled limit (\ref{DS}) that could access such small effects. 

\textbf{2.} Another finite-$N$ effect that is missing from the collective field description is the $N \mod 8 $ dependence of the exact spectral form factor, which is related to the particle-hole symmetry \cite{You16,Cotler16,Garcia-Garcia16}.

\textbf{3.} The Thouless time that we have obtained scales like $\sqrt{N} \log N$ with $N$, which agrees with the analytic results of \cite{Altland17} but not the numerical results of \cite{Gharibyan18}. We speculate that in our case the Thouless time result can be made more precise by considering higher orders of the $1/q$-expansion and its appropriate extrapolation. 

\textbf{4.} To obtain the ramp solution, we have used the approach of \cite{Maldacena18}, used to construct the solution in the system of two coupled SYK chains that is dual to a Euclidean wormhole. The similarities on the SYK side are in agreement with the similarities between the eternal wormhole and the Lorentzian double cone geometry of \cite{Saad18} on the gravity side. However, we have also found that on the slope regime there are subleading saddles. There are no such saddles on the ramp, but there are replica-nondiagonal analogues at finite $q$ for the Euclidean replica partition function \cite{AKTV,AKV}. This provides a new constraint on the question of importance of the subleading SYK saddles in the gravity description.

\section*{Acknowledgements}

The authors are grateful to Irina Aref'eva, Changha Choi, Mark Mezei and Gabor Sarosi for useful discussions, to Irina Aref'eva for comments on the draft and to Boris Eremin for collaboration on the early stage of the work. The work of M. K., which consisted in designing and directing the project and interpreting the results is supported by the Russian Science foundation under grant no.~20-12-00200 and performed in Steklov Mathematical Institute of Russian Academy of Sciences.

\appendix
\section{Spectral form factor in SYK as a collective field path integral}

\label{sec:ReplicasDeriv}

In this section we review the derivation of the disorder-averaged replica path integral (\ref{S(T)-bilocal}). We start from the product of the two fermionic path integrals  (\ref{Z(iT)}) and (\ref{Z(-iT)}):
\bea
Z(iT) Z(-i T) &=& \int D\psi^L D \psi^R  \exp\left[  \int_0^T dt \left(-\frac{1}{2} \psi_i^L \dd_t \psi_i^L - \frac{1}{2} \psi_i^R \dd_t \psi_i^R \right.\right.\nn\\&-& \left.\left.\left(i^{q/2}\!\!\!\!\! \sum_{i_1< \dots < i_q=1}^N \!\!\!\!\! j_{i_1 i_2\dots i_q} \psi_{i_1} \dots \psi_{i_q} - (-i)^{q/2}\!\!\!\!\! \sum_{i_1< \dots < i_q=1}^N \!\!\!\!\! j_{i_1 i_2\dots i_q} \psi_{i_1} \dots \psi_{i_q}\right)\right)\right],
\eea
where we assume a fixed realization of the randomized couplings  ${\bf j} = \{j_{i_1\dotsi_q}\}$. As a first step, we integrate out the disorder ${\bf j}$ with the measure (\ref{Gaussian}). As a result, we arrive at the disorder-averaged fermionic path integral: 
\bea\label{Sreplica-in}
&& S(T) = \int D\psi^\alpha \exp\left[-\frac12 \sum_{\alpha=L,R} \int dt \left(\psi^\alpha_i \dd_t \psi_i^\alpha\right) - \frac{2^{q-2} \mJ^2 }{q^2 N^{q-1}} i^q \right.\\\nn &&\times \left.\int dt \left(\psi_{i_1}^L(t) \dots \psi_{i_q}^L (t) - (-1)^{\frac{q}{2}}\psi_{i_1}^R(t) \dots \psi_{i_q}^R (t)\right) \int dt'\left(\psi_{i_1}^L(t') \dots \psi_{i_q}^L (t') - (-1)^{\frac{q}{2}}\psi_{i_1}^R(t') \dots \psi_{i_q}^R (t')\right) \right]\,. 
\eea
Here and henceforth the unordered sum over all repeating color indices is implicit. Let us now introduce the notations for the constant matrix $s_{\alpha\beta}$ and a bilinear combination $\Xi_{\alpha\beta} (t, t')$ (following alongside the partition function derivation in \cite{Kitaev17}): 
\bea
&& s_{LL} = s_{RR} = -1\,; \quad s_{LR} = s_{RL} = (-1)^{\frac{q}{2}} = i^q \,;\\
&& \Xi_{\alpha\beta}(\tau, \tau') = -\frac{1}{N} \psi_k^\alpha (\tau) \psi_k^\beta(\tau')\,. \label{Xi}
\eea
Anticommuting the fermions in (\ref{Sreplica-in}) into bilinears, we can use these notations to rewrite (\ref{Sreplica-in}) as follows: 
\be\label{Sreplica-in-2}
S(T) = \int D\psi^\alpha \exp\left[-\frac12 \int dt \left(\psi^\alpha_i \dd_t \psi_i^\alpha\right) + \frac{2^{q-2} N \mJ^2 }{q^2 }\int dt \int dt's_{\alpha\beta} \Xi_{\alpha\beta} (t, t')^q \right]\,. 
\ee
One can formally express any functional $F[\Xi]$ by using a delta function: 
\bea
F[\Xi] &=& \int DG_{\alpha\beta}(\tau, \tau')\ F[-G]\ \delta(G_{\alpha\beta}(\tau, \tau') + \Xi_{\alpha\beta}(\tau, \tau'))\label{Identity}\\ &=& \int D G_{\alpha\beta}(\tau, \tau') \ F[-G] \int D\Sigma_{\alpha\beta}(\tau, \tau')\ \e^{-\frac{N}{2}\int d\tau_1 d\tau_2 \Sigma_{\alpha\beta}(\tau_1, \tau_2) [G_{\alpha\beta}(\tau_1, \tau_2) + \Xi_{\alpha\beta}(\tau_1, \tau_2)]}  \,.\nn 
\eea
In the last step the generalized Laplace transform was applied to the delta-function. Using this representation, we write 
\bea
F[\Xi] &=& \exp \left( \frac{2^{q-2} N \mJ^2 }{q^2 } \int\int d\tau d\tau' s_{\alpha\beta} \Xi_{\alpha \beta} (\tau, \tau')^q \right)\\ &=& \nn \int DG D\Sigma\ \exp\left[\frac{N}{2} \int\int d\tau d\tau' \left(\frac{2^{q-1} \mJ^2}{q^2} s_{\alpha\beta} G_{\alpha\beta}(\tau, \tau')^q - \Sigma_{\alpha\beta} (\tau, \tau') (G_{\alpha\beta}(\tau, \tau') + \Xi_{\alpha\beta}(\tau, \tau'))\right)\right].
\eea
Substituting into (\ref{Sreplica-in-2}) and integrating out the fermions, one arrives at the resulting expression (\ref{S(T)-bilocal}):
\be
S(T) = \int DG_{\alpha\beta} D \Sigma_{\alpha\beta} \e^{-N I[G, \Sigma]}\,, 
\ee
where the action has the form \cite{Saad18}:
\bea
I[G, \Sigma] 
=&&  -\log \Pf[\delta_{\alpha\beta} \dd_t - \hat{\Sigma}_{\alpha\beta}] + \\&&\nn \frac{1}{2}\int_0^T \int_0^T dt_1 dt_2  \left(\Sigma_{\alpha\beta}(t_1, t_2) G_{\alpha\beta}(t_1, t_2) - \frac{2^{q-1} \mJ^2}{q^2}s_{\alpha\beta} G_{\alpha\beta}(t_1, t_2)^q\right)\,.
\eea
The path integral over the bilocal fields $G$ and $\Sigma$ is assumed to be taken over an appropriate complex contour which ensures convergence.

\section{One-loop determinant for off-diagonal modes on the slope}
\label{sec:1loop-slope-offdiag}

In this appendix we compute the inverse determinant of a free differential operator on the space of anti-periodic bilocal fields which appears in (\ref{SFFslope}):
\be
\Delta_{\text{free}} = \det\left( \dd_u^2 - \frac14 \dd_v^2\right)^{-1}\,.
\ee
The eigenfunctions of the operator $\left(\partial_{u}^2-\frac{1}{4}\partial^2_{v}\right)$ are
\begin{equation}
	\mathcal{F}(u,v)=\e^{i\omega u}\e^{i\rho v}.
\end{equation}
These functions must be antiperiodic by $t_1$ and $t_2$ with the period $T$, so
\begin{equation}
	\mathcal{F}\left(u+\mJ T, v\pm\frac{\mJ T}{2}\right)=-\mathcal{F}(u, v), 
\end{equation}
and therefore
\begin{equation}
	\mathcal{F}(u ,v)=\begin{cases}\e^{i\pi k \frac{u}{\mJ T}}\e^{2i\pi n \frac{v}{\mJ T}},\quad n\in2\mathbb{Z},\quad k\in2\mathbb{Z}+1, \\
		\e^{i\pi k \frac{u}{\mJ T}}\e^{2i\pi n \frac{v}{\mJ T}},\quad n\in2\mathbb{Z}+1,\quad k\in2\mathbb{Z}\,.
	\end{cases}
\end{equation}
So the eigenvalues are
\begin{equation}
	\left(\partial_{u}^2-\frac{1}{4}\partial^2_{v}\right)\mathcal{F}(u, v)=\frac{\pi^2}{(\mJ T)^2}(n^2-k^2)\mathcal{F}(u,v).
\end{equation}
Thus the inverse determinant reads
\begin{equation}
	\begin{aligned}
		\frac{1}{\det\frac{1}{\mJ^2}\left(\partial_{t}^2-\frac{1}{4}\partial^2_{T'}\right)}&=\prod\limits_{\substack{n\in 2\mathbb{Z}\\ k\in 2\mathbb{Z}+1}}\prod\limits_{\substack{n\in 2\mathbb{Z}+1\\ k\in 2\mathbb{Z}}}\left(\frac{\mJ T}{\pi }\right)^2\frac{1}{n^2-k^2}=\prod\limits_{k\in2\mathbb{Z}+1}\prod\limits_{n\in2\mathbb{Z}+1}\left(\frac{\mJ T}{\pi }\right)^2\frac{1}{nk}=\\
		&=\prod\limits_{n\in (2\mathbb{Z}+1)_+}\left(\frac{\mJ T}{\pi n}\right)^4=\left(\frac{\mJ T}{\pi}\right)^{4\sum\limits_{n\in (2\mathbb{Z}+1)_+}1}\e^{-4\sum\limits_{n=1}^{+\infty}\log(2n-1)}=\frac{1}{4},
	\end{aligned}
\end{equation}
as we dropped out singular terms $\Lambda_1$ and $\Lambda_2$ using zeta-function regularization:
\begin{equation}
	(1-2^{-s})\zeta(s)\Big|_{s=0}=\sum\limits_{n\in (2\mathbb{Z}+1)_+}1=\Lambda_1,\text{ and }-\sum_{n=1}^{+\infty}\log(2n-1)=\zeta(0)\log2=-\frac{\log 2}{2}+\Lambda_2\,.
\end{equation}

\section{One-loop effective action on the ramp}
\label{sec:RampQuantum-deriv}

Here we derive the $1$-loop quantum action for the spectral form factor on the background of the replica-nondiagonal solution. We start from the action~\eqref{action} and split bilocal replica fields into classical and quantum parts:
\begin{equation}
	G_{\alpha\beta}=G^{\text{cl}}_{\alpha\beta}+\mathfrak{G}_{\alpha\beta},\quad \Sigma_{\alpha\beta}=\Sigma^{\text{cl}}_{\alpha\beta}+\mathfrak{s}_{\alpha\beta}.
\end{equation}
The quantum part of the action is:
\begin{equation}
	\begin{aligned}
		I_q&=-\frac{1}{4J^2(q-1)}\int dt_1 dt_2 dt_3 dt_4\,\mathfrak{s}_{\alpha\beta}(t_1,t_2)K_{\alpha\beta\gamma\delta}(t_1,t_2;t_3,t_4)\mathfrak{s}_{\gamma\delta}(t_3,t_4)+\\
		&+\frac{1}{2}\int dt_1 dt_2\left[\mathfrak{G}_{\alpha\beta}(t_1,t_2)\mathfrak{s}_{\alpha\beta}(t_1,t_2)-\frac{1}{2}J^2(q-1)G^{\text{cl}}_{\alpha\beta}(t_1,t_2)^{q-2}\mathfrak{G}^2_{\alpha\beta}(t_1,t_2)\right].
	\end{aligned}
\end{equation}
Here we introduce the ladder kernel $K$ in analogy to Euclidean SYK partition function \cite{MScomments}:
\begin{equation}
	K_{\alpha\beta\gamma\delta}(t_1,t_2;t_3,t_4)=-J^2(q-1)G^{\text{cl}}_{\alpha\gamma}(t_1-t_3)G^{\text{cl}}_{\beta\delta}(t_2-t_4).
\end{equation}
Integrating out $\mathfrak{s}_{\alpha\beta}$
\begin{equation}
	\mathfrak{s}=J^2(q-1)K^{-1}*\mathfrak{G},
\end{equation}
we get:
\bea\nn
I_q &=& \frac{J^2(q-1)}{4}\left[\int dt_1 dt_2 dt_3 dt_4\,\mathfrak{G}_{\alpha\beta}(t_1,t_2)K^{-1}_{\alpha\beta\gamma\delta}(t_1,t_2;t_3,t_4)\mathfrak{G}_{\gamma\delta}(t_3,t_4) \right.\\ &-& \left.\int dt_1 dt_2\,G^{\text{cl}}_{\alpha\beta}(t_1-t_2)^{q-2} s_{\alpha\beta}	\mathfrak{G}^2_{\alpha\beta}(t_1,t_2) \right]\,.
\eea
The spectral form factor path integral gives
\begin{equation}
	\begin{aligned}
		S(T)&\to \int\mathcal{D}\mathfrak{G}_{\alpha\beta}\exp\bigg\{\frac{N}{4}\bigg[\frac{1}{4}\int dt_1 dt_2dt_3 dt_4\,\mathfrak{G}_{\alpha\beta}(t_1,t_2)\widetilde{K}^{-1}_{\alpha\beta\gamma\delta}(t_1,t_2;t_3,t_4)\mathfrak{G}_{\gamma\delta}(t_3,t_4)+\\
		&+\frac{\mJ^2}{2}\int du dv\,\left[2G^{\text{cl}}_{\alpha\beta}(u)\right]^{q}\left(G^{\text{cl}}_{\alpha\beta}(u)\right)^{-2}s_{\alpha\beta}\mathfrak{G}^2_{\alpha\beta}(u,v)\bigg]\bigg\},
	\end{aligned}
\end{equation}
where 
\begin{equation}
	\widetilde{K}_{\alpha\beta\gamma\delta}(t_1,t_2;t_3,t_4)=G^{\text{cl}}_{\alpha\gamma}(t_1-t_3)G^{\text{cl}}_{\beta\delta}(t_2-t_4).
\end{equation}
So we need to find $(G_{\alpha\beta}^{\text{cl}}(t_1-t_2))^{-1}$, which is defined by
\begin{equation}
	\int dt (G_{\alpha\gamma}^{\text{cl}}(t_1-t))^{-1}G_{\gamma\beta}^{\text{cl}}(t-t_2)=\delta(t_1-t_2)\delta_{\alpha\beta}.
\end{equation}
Let us check that $(G_{\alpha\gamma}^{\text{cl}}(t_1-t_3))^{-1}(G_{\beta\delta}^{\text{cl}}(t_2-t_4))^{-1}=\widetilde{K}^{-1}_{\alpha\beta\gamma\delta}(t_1,t_2;t_3,t_4)$:
\begin{equation}
	\int dt dt'(G_{\alpha\gamma'}^{\text{cl}}(t_1-t))^{-1}(G_{\beta\delta'}^{\text{cl}}(t_2-t'))^{-1}G^{\text{cl}}_{\gamma'\gamma}(t-t_3)G^{\text{cl}}_{\delta'\delta}(t'-t_4)=\delta(t_1-t_3)\delta_{\alpha\gamma}\delta(t_2-t_4)\delta_{\beta\delta}.
\end{equation}
For small times we can omit $1/q$ terms in Green functions in $\widetilde{K}_{\alpha\beta\gamma\delta}$ and get that
\begin{equation}
	(G_{\alpha\beta}^{\text{cl}}(t_1-t_2))^{-1}=\delta'(t_1-t_2)\delta_{\alpha\beta}.
\end{equation}
For big times we can use the equation of motion for the Green function~\eqref{EOM}:
\be 
\int dt\left[\delta(t-t_1)\delta_{\alpha\gamma}\partial_{t}-\Sigma^{\text{cl}}_{\alpha\gamma}(t_1,t)\right]G^{\text{cl}}_{\gamma\beta}(t,t_2)=\delta(t_1-t_2)\delta_{\alpha\beta},
\ee
from which it follows that:
\begin{equation}
	(G_{\alpha\beta}^{\text{cl}}(t_1-t_2))^{-1}=\delta'(t_1-t_2)\delta_{\alpha\beta}-\Sigma^{\text{cl}}_{\alpha\beta}(t_1,t_2).
\end{equation}
Here for large times we can drop out $\Sigma^{\text{cl}}_{\alpha\beta}$, as it is zero or proportional to delta-function. acting outside the integration interval.  

Thus, for all time ranges we get the following expression for the $1$-loop correction to the spectral form factor in the large $q$ limit:
\begin{equation}
	\begin{aligned}
		S(T)&=\int\mathcal{D}\mathfrak{G}_{\alpha\beta}\exp\bigg\{\frac{N}{8}\bigg[\frac{1}{2}\int du dv\,\mathfrak{G}_{\alpha\beta}(u,v)\left(\frac{1}{4}\partial^2_{v}-\partial^2_{u}\right)\mathfrak{G}_{\alpha\beta}(u,v)+\\
		&+\mJ^2\int du dv\,\left[2G^{\text{cl}}_{\alpha\beta}(u)\right]^{q}\left(G^{\text{cl}}_{\alpha\beta}(u)\right)^{-2}s_{\alpha\beta}\mathfrak{G}^2_{\alpha\beta}(u,v)\bigg]\bigg\}.
	\end{aligned} \label{1loop-ramp-full}
\end{equation}
This expression coincides with~\eqref{SFFramp} for small times if we identify
\begin{equation}
	\mathfrak{G}_{\alpha\beta}=\frac{1}{q}\mathfrak{g}_{\alpha\beta},
\end{equation}
and is smoothly extended to the late time region, as we glued the Green functions. Also note that for large times the potential term vanishes, as it is proportional to $\Sigma^{\text{cl}}_{\alpha\beta}$ which is zero or delta-function acting outside the integration interval. Also, the potential term in~\eqref{SFFramp} exponentially decays for large times. Therefore, we can use small time solutions on the entire integration interval.




\begin{thebibliography}{99}
\bibitem{Sachdev92} 
  S.~Sachdev and J.~Ye,
  ``Gapless spin fluid ground state in a random, quantum Heisenberg magnet,''
  Phys.\ Rev.\ Lett.\  {\bf 70}, 3339 (1993)
  doi:10.1103/PhysRevLett.70.3339
  [cond-mat/9212030].

\bibitem{Kitaev} A.~Kitaev, talks at KITP in 2015: \href{http://online.kitp.ucsb.edu/online/entangled15/kitaev/}{http://online.kitp.ucsb.edu/online/entangled15/kitaev/}, \href{http://online.kitp.ucsb.edu/online/entangled15/kitaev2/}{http://online.kitp.ucsb.edu/online/entangled15/kitaev2/}

\bibitem{MScomments} 
  J.~Maldacena and D.~Stanford,
  ``Remarks on the Sachdev-Ye-Kitaev model,''
  Phys.\ Rev.\ D {\bf 94}, no. 10, 106002 (2016)
  [arXiv:1604.07818 [hep-th]].

\bibitem{Kitaev17} 
  A.~Kitaev and S.~J.~Suh,
  ``The soft mode in the Sachdev-Ye-Kitaev model and its gravity dual,''
  arXiv:1711.08467 [hep-th].

\bibitem{Jensen16} 
K.~Jensen,
``Chaos in AdS$_2$ Holography,''
Phys.\ Rev.\ Lett.\  {\bf 117}, no. 11, 111601 (2016)
[arXiv:1605.06098 [hep-th]].

\bibitem{Maldacena16} 
J.~Maldacena, D.~Stanford and Z.~Yang,
``Conformal symmetry and its breaking in two dimensional Nearly Anti-de-Sitter space,''
PTEP {\bf 2016}, no. 12, 12C104 (2016)
doi:10.1093/ptep/ptw124
[arXiv:1606.01857 [hep-th]].

\bibitem{Engelsoy16} 
J.~Engelsöy, T.~G.~Mertens and H.~Verlinde,
``An investigation of AdS$_{2}$ backreaction and holography,''
JHEP {\bf 1607}, 139 (2016)
doi:10.1007/JHEP07(2016)139
[arXiv:1606.03438 [hep-th]].

\bibitem{Jevicki16} 
A.~Jevicki, K.~Suzuki and J.~Yoon,
``Bi-Local Holography in the SYK Model,''
JHEP {\bf 1607}, 007 (2016)
doi:10.1007/JHEP07(2016)007
[arXiv:1603.06246 [hep-th]].

\bibitem{Cotler16} 
  J.~S.~Cotler {\it et al.},
  ``Black Holes and Random Matrices,''
  JHEP {\bf 1705}, 118 (2017)
  doi:10.1007/JHEP05(2017)118
  [arXiv:1611.04650 [hep-th]].
  
\bibitem{Maldacena18} 
X.~Maldacena and X.~L.~Qi,
``Eternal traversable wormhole,''
arXiv:1804.00491 [hep-th].

\bibitem{Harlow18}
D.~Harlow and D.~Jafferis,
``The Factorization Problem in Jackiw-Teitelboim Gravity,''
JHEP \textbf{02}, 177 (2020)
[arXiv:1804.01081 [hep-th]].

\bibitem{Saad18} 
P.~Saad, S.~H.~Shenker and D.~Stanford,
``A semiclassical ramp in SYK and in gravity,''
arXiv:1806.06840 [hep-th].
  
\bibitem{Goel18}
A.~Goel, H.~T.~Lam, G.~J.~Turiaci and H.~Verlinde,
``Expanding the Black Hole Interior: Partially Entangled Thermal States in SYK,''
JHEP \textbf{02}, 156 (2019)
[arXiv:1807.03916 [hep-th]].
  
\bibitem{Garcia-Garcia19}
A.~M.~Garc\'\i{}a-Garc\'\i{}a, T.~Nosaka, D.~Rosa and J.~J.~M.~Verbaarschot,
``Quantum chaos transition in a two-site Sachdev-Ye-Kitaev model dual to an eternal traversable wormhole,''
Phys. Rev. D \textbf{100}, no.2, 026002 (2019)
[arXiv:1901.06031 [hep-th]].
  
\bibitem{SSS}
P.~Saad, S.~H.~Shenker and D.~Stanford,
``JT gravity as a matrix integral,''
[arXiv:1903.11115 [hep-th]].

\bibitem{Saad19}
P.~Saad,
``Late Time Correlation Functions, Baby Universes, and ETH in JT Gravity,''
[arXiv:1910.10311 [hep-th]].

\bibitem{Gao19}
P.~Gao and D.~L.~Jafferis,
``A Traversable Wormhole Teleportation Protocol in the SYK Model,''
[arXiv:1911.07416 [hep-th]].
  
\bibitem{Garcia-Garcia20}
A.~M.~Garc\'\i{}a-Garc\'\i{}a, Y.~Jia, D.~Rosa and J.~J.~M.~Verbaarschot,
``Sparse Sachdev-Ye-Kitaev model, quantum chaos and gravity duals,''
[arXiv:2007.13837 [hep-th]].  

\bibitem{You16}
Y.~Z.~You, A.~W.~W.~Ludwig and C.~Xu,
``Sachdev-Ye-Kitaev Model and Thermalization on the Boundary of Many-Body Localized Fermionic Symmetry Protected Topological States,''
Phys. Rev. B \textbf{95}, no.11, 115150 (2017)
[arXiv:1602.06964 [cond-mat.str-el]].

\bibitem{Garcia-Garcia16}
A.~M.~Garc\'\i{}a-Garc\'\i{}a and J.~J.~M.~Verbaarschot,
``Spectral and thermodynamic properties of the Sachdev-Ye-Kitaev model,''
Phys. Rev. D \textbf{94}, no.12, 126010 (2016)
[arXiv:1610.03816 [hep-th]].
 
\bibitem{Li17}
T.~Li, J.~Liu, Y.~Xin and Y.~Zhou,
``Supersymmetric SYK model and random matrix theory,''
JHEP \textbf{06}, 111 (2017)
doi:10.1007/JHEP06(2017)111
[arXiv:1702.01738 [hep-th]].
 
\bibitem{delCampo17}
A.~del Campo, J.~Molina-Vilaplana and J.~Sonner,
``Scrambling the spectral form factor: unitarity constraints and exact results,''
Phys. Rev. D \textbf{95}, no.12, 126008 (2017)
[arXiv:1702.04350 [hep-th]].

\bibitem{Sonner17}
J.~Sonner and M.~Vielma,
``Eigenstate thermalization in the Sachdev-Ye-Kitaev model,''
JHEP \textbf{11}, 149 (2017)
[arXiv:1707.08013 [hep-th]].
 
\bibitem{Hunter-Jones17}
N.~Hunter-Jones and J.~Liu,
``Chaos and random matrices in supersymmetric SYK,''
JHEP \textbf{05}, 202 (2018)
[arXiv:1710.08184 [hep-th]].
 
\bibitem{Altland17}
A.~Altland and D.~Bagrets,
``Quantum ergodicity in the SYK model,''
Nucl. Phys. B \textbf{930}, 45-68 (2018)
[arXiv:1712.05073 [cond-mat.str-el]].
 
\bibitem{Garcia-Garcia18}
A.~M.~Garc\'\i{}a-Garc\'\i{}a, Y.~Jia and J.~J.~M.~Verbaarschot,
``Universality and Thouless energy in the supersymmetric Sachdev-Ye-Kitaev Model,''
Phys. Rev. D \textbf{97}, no.10, 106003 (2018)
[arXiv:1801.01071 [hep-th]]. 
 
\bibitem{Roberts18}
D.~A.~Roberts, D.~Stanford and A.~Streicher,
``Operator growth in the SYK model,''
JHEP \textbf{06}, 122 (2018)
[arXiv:1802.02633 [hep-th]].
 
\bibitem{Gharibyan18}
H.~Gharibyan, M.~Hanada, S.~H.~Shenker and M.~Tezuka,
``Onset of Random Matrix Behavior in Scrambling Systems,''
JHEP \textbf{07}, 124 (2018)
[erratum: JHEP \textbf{02}, 197 (2019)]
[arXiv:1803.08050 [hep-th]].
 
\bibitem{Qi18} 
X.~L.~Qi and A.~Streicher,
``Quantum Epidemiology: Operator Growth, Thermal Effects, and SYK,''
JHEP {\bf 1908}, 012 (2019)
[arXiv:1810.11958 [hep-th]].

\bibitem{Lau18}
P.~H.~C.~Lau, C.~T.~Ma, J.~Murugan and M.~Tezuka,
``Randomness and Chaos in Qubit Models,''
Phys. Lett. B \textbf{795}, 230-235 (2019)
[arXiv:1812.04770 [hep-th]].

\bibitem{Nayak19}
P.~Nayak, J.~Sonner and M.~Vielma,
``Extended Eigenstate Thermalization and the role of FZZT branes in the Schwarzian theory,''
JHEP \textbf{03}, 168 (2020)
[arXiv:1907.10061 [hep-th]]. 

\bibitem{Lucas19}
A.~Lucas,
``Non-perturbative dynamics of the operator size distribution in the Sachdev\textendash{}Ye\textendash{}Kitaev model,''
J. Math. Phys. \textbf{61}, no.8, 081901 (2020)
[arXiv:1910.09539 [hep-th]].

\bibitem{Jia19}
Y.~Jia and J.~J.~M.~Verbaarschot,
``Spectral Fluctuations in the Sachdev-Ye-Kitaev Model,''
JHEP \textbf{07}, 193 (2020)
[arXiv:1912.11923 [hep-th]].

\bibitem{Winer20}
M.~Winer, S.~K.~Jian and B.~Swingle,
``An exponential ramp in the quadratic Sachdev-Ye-Kitaev model,''
[arXiv:2006.15152 [cond-mat.stat-mech]].

\bibitem{Altland20}
A.~Altland and J.~Sonner,
``Late time physics of holographic quantum chaos,''
[arXiv:2008.02271 [hep-th]].

\bibitem{Choi20}
C.~Choi, M.~Mezei and G.~S\'arosi,
``Pole skipping away from maximal chaos,''
[arXiv:2010.08558 [hep-th]].

\bibitem{Cardella19}
M.~A.~Cardella,
``Derivation of the two Schwarzians effective action for the Sachdev-Ye-Kitaev spectral form factor,''
[arXiv:1907.09570 [hep-th]].

\bibitem{Okuyama19}
K.~Okuyama,
``Replica symmetry breaking in random matrix model: a toy model of wormhole networks,''
Phys. Lett. B \textbf{803}, 135280 (2020)
[arXiv:1903.11776 [hep-th]].

\bibitem{Arefeva192}
I.~Aref'eva and I.~Volovich,
``Gas of baby universes in JT gravity and matrix models,''
Symmetry \textbf{12}, no.6, 975 (2020)
[arXiv:1905.08207 [hep-th]].

\bibitem{Stanford19}
D.~Stanford and E.~Witten,
``JT Gravity and the Ensembles of Random Matrix Theory,''
[arXiv:1907.03363 [hep-th]].

\bibitem{Johnson19}
C.~V.~Johnson,
``Nonperturbative Jackiw-Teitelboim gravity,''
Phys. Rev. D \textbf{101}, no.10, 106023 (2020)
[arXiv:1912.03637 [hep-th]].

\bibitem{Penington191}
G.~Penington,
``Entanglement Wedge Reconstruction and the Information Paradox,''
JHEP \textbf{09}, 002 (2020)
[arXiv:1905.08255 [hep-th]].

\bibitem{Almheiri191} 
A.~Almheiri, N.~Engelhardt, D.~Marolf and H.~Maxfield,
``The entropy of bulk quantum fields and the entanglement wedge of an evaporating black hole,''
JHEP {\bf 1912}, 063 (2019)
[arXiv:1905.08762 [hep-th]].

\bibitem{Almheiri192} 
A.~Almheiri, R.~Mahajan, J.~Maldacena and Y.~Zhao,
``The Page curve of Hawking radiation from semiclassical geometry,''
JHEP \textbf{03}, 149 (2020)
[arXiv:1908.10996 [hep-th]].

\bibitem{Penington192} 
G.~Penington, S.~H.~Shenker, D.~Stanford and Z.~Yang,
``Replica wormholes and the black hole interior,''
arXiv:1911.11977 [hep-th].

\bibitem{Almheiri194} 
A.~Almheiri, T.~Hartman, J.~Maldacena, E.~Shaghoulian and A.~Tajdini,
``Replica Wormholes and the Entropy of Hawking Radiation,''
JHEP \textbf{05}, 013 (2020)
[arXiv:1911.12333 [hep-th]].

\bibitem{Gur-Ari18}
G.~Gur-Ari, R.~Mahajan and A.~Vaezi,
``Does the SYK model have a spin glass phase?,''
JHEP \textbf{11}, 070 (2018)
[arXiv:1806.10145 [hep-th]].

\bibitem{AKTV}
I.~Aref'eva, M.~Khramtsov, M.~Tikhanovskaya and I.~Volovich,
``Replica-nondiagonal solutions in the SYK model,''
JHEP \textbf{07}, 113 (2019)
[arXiv:1811.04831 [hep-th]].

\bibitem{Wang18}
H.~Wang, D.~Bagrets, A.~L.~Chudnovskiy and A.~Kamenev,
``On the replica structure of Sachdev-Ye-Kitaev model,''
JHEP \textbf{09}, 057 (2019)
[arXiv:1812.02666 [hep-th]].

\bibitem{Arefeva19}
I.~Aref'eva and I.~Volovich,
``Spontaneous symmetry breaking in fermionic random matrix model,''
JHEP \textbf{10}, 114 (2019)
[arXiv:1902.09970 [hep-th]].

\bibitem{AKV}
I.~Ya.~Aref'eva, I.~Volovich and M.~A.~Khramtsov,
``Revealing nonperturbative effects in the SYK model,''
Theor. Math. Phys. \textbf{201}, no.2, 1585-1605 (2019)
[arXiv:1905.04203 [hep-th]].
  
\bibitem{Schroder}
L.~Erdős and D.~Schröder, ``Phase Transition in the Density of States of Quantum Spin Glasses,'' Math Phys Anal Geom 17, 441–464 (2014).

\bibitem{Eberlein17}
A.~Eberlein, V.~Kasper, S.~Sachdev and J.~Steinberg,
``Quantum quench of the Sachdev-Ye-Kitaev Model,''
Phys. Rev. B \textbf{96}, no.20, 205123 (2017)
[arXiv:1706.07803 [cond-mat.str-el]].

\bibitem{Bhattacharya17}
R.~Bhattacharya, S.~Chakrabarti, D.~P.~Jatkar and A.~Kundu,
``SYK Model, Chaos and Conserved Charge,''
JHEP \textbf{11}, 180 (2017)
[arXiv:1709.07613 [hep-th]].

\bibitem{Tarnopolsky18}
G.~Tarnopolsky,
``Large $q$ expansion in the Sachdev-Ye-Kitaev model,''
Phys. Rev. D \textbf{99}, no.2, 026010 (2019)
[arXiv:1801.06871 [hep-th]].

\bibitem{Berkooz18} 
M.~Berkooz, M.~Isachenkov, V.~Narovlansky and G.~Torrents,
``Towards a full solution of the large N double-scaled SYK model,''
JHEP {\bf 1903}, 079 (2019)
[arXiv:1811.02584 [hep-th]].
  
\bibitem{Jiang19}
J.~Jiang and Z.~Yang,
``Thermodynamics and Many Body Chaos for generalized large q SYK models,''
JHEP \textbf{08}, 019 (2019)
[arXiv:1905.00811 [hep-th]].

\bibitem{Streicher19}
A.~Streicher,
``SYK Correlators for All Energies,''
JHEP \textbf{02}, 048 (2020)
[arXiv:1911.10171 [hep-th]].

\bibitem{Choi19}
C.~Choi, M.~Mezei and G.~S\'arosi,
``Exact four point function for large $q$ SYK from Regge theory,''
[arXiv:1912.00004 [hep-th]].

\bibitem{Nosaka19}
T.~Nosaka and T.~Numasawa,
``Quantum Chaos, Thermodynamics and Black Hole Microstates in the mass deformed SYK model,''
JHEP \textbf{08}, 081 (2020)
[arXiv:1912.12302 [hep-th]].

\bibitem{Berkooz20}
M.~Berkooz, N.~Brukner, V.~Narovlansky and A.~Raz,
``The double scaled limit of Super--Symmetric SYK models,''
[arXiv:2003.04405 [hep-th]].

\bibitem{Das20}
S.~R.~Das, A.~Ghosh, A.~Jevicki and K.~Suzuki,
``Near Conformal Perturbation Theory in SYK Type Models,''
[arXiv:2006.13149 [hep-th]].

\bibitem{Nosaka18}
T.~Nosaka, D.~Rosa and J.~Yoon,
``The Thouless time for mass-deformed SYK,''
JHEP \textbf{09}, 041 (2018)
[arXiv:1804.09934 [hep-th]].

\bibitem{Lekner}
J.~Lekner,``Reflectionless eigenstates of the sech 2 potential,''
American Journal of Physics \textbf{75} no. 12 (2007) 1151--1157

\end{thebibliography}
\end{document}